\documentclass{article}
\usepackage[utf8]{inputenc}
\usepackage{natbib}
\usepackage{xcolor}
\usepackage{url}
\usepackage{alphabeta}
\usepackage{amssymb}
\usepackage{amsmath,bm,calc}
\usepackage{hyperref}
\usepackage{subfigure}
\usepackage{graphicx}
\hypersetup{
    colorlinks=true,
    linkcolor=blue,
    citecolor=blue
    } 
\usepackage[hmarginratio=1:1]{geometry}
\usepackage{cleveref}
\usepackage{comment} 
\usepackage{float}
\usepackage{tabularx} 
\usepackage{algorithm}
\usepackage{algpseudocode}

\newcommand{\logit}{\mbox{logit}}
\usepackage{authblk}

\title{Modelling multivariate ordinal time series using pairwise likelihood}

\author{Anna Nalpantidi}
\author{Dimitris Karlis}

\affil{Department of Statistics, \\ Athens University of Economics and Business,  Greece \\ 
\texttt{analpa@aueb.gr}}

\date{February 2026}

\begin{document}

\maketitle

\begin{abstract}
We assume that we have multiple ordinal time series and we would like to specify their joint distribution. In general it is difficult to create  multivariate distribution that can be easily used to jointly model ordinal variables and the problem becomes even more complex in the case of time series, since we have to take into consideration not only the autocorrelation of each time series and the dependence between time series, but also cross-correlation.
Starting from the simplest case of two ordinal time series, we propose using copulas to specify their joint distribution. We extend our approach in higher dimensions, by approximating full likelihood with composite likelihood and especially conditional pairwise likelihood, where each bivariate model is specified by copulas. We suggest maximizing each bivariate model independently to avoid computational issues and synthesize individual  estimates using weighted mean. Weights are related to the Hessian matrix of each bivariate model.  Simulation studies showed that model fits well under different sample sizes. Forecasting approach is also discussed. A small real data application about unemployment state of different countries of European Union is presented to illustrate our approach.
\end{abstract}


\section{Introduction}

We consider the case of time series when the observations are ordinal random variables. 
More specifically, we assume that the time series $Z_t$, $t=1,\ldots,T$ takes values in a finite number of $d$ categories/states $\mathcal{S}=\{s_1,s_2,\ldots, s_d\}$, $Z_t \in S$. Its characteristic is that categories present a natural ordering so we assume that $s_1<s_2<\ldots<s_d$.

Ordinal time series have found several applications, including
 environmental problems \citep{liu2022modeling1, jahn2024nonlinear}, health studies \citep{fokianos2003regression} and  economics \citep{weiss2019distance}, among others.
  There have been several  methods proposed for modeling ordinal time series in the literature. Especially, as univariate models for ordinal time series are concerned, there is a great variety of approaches like Markov models \citep{raftery1985model,buhlmann1999variable}, regression models \citep{gottlein1992ordinal, pruscha1993categorical, fokianos2003regression}, latent continuous models \citep{varin2006pairwise}, GARCH-type models \citep{weiss2023ordinal} or methods that treat ordinal variables as rank-count variables \citep{weiss2019distance,liu2022modeling2}.

On the other hand, the literature is still limited about  models for multivariate ordinal time series. There are only  few works for longitudinal/panel ordinal data and time series in the literature. More particularly, \cite{nikoloulopoulos2016copula}, and \cite{nikoloulopoulos2019coupling} proposed modeling each ordinal time series through a copula-based Markov model and in case of multiple ordinal time series we can jointly model them by assuming an appropriate multivariate copula. 

In a different approach, \cite{chaubert2008multivariate} developed a Dynamic Multivariate Probit Ordinal Model (DMPOM), as part of Generalized Linear Multivariate Mixed Models (GLMMM), to model multiple ordinal variables capturing both forms of correlation: between variables and across each variable. Also working with the latent continuous variable, \cite{hirk2019multivariate} assumed a linear model for the latent variable of each subject at a specific time, for a specific outcome. Related works can be seen in  \cite{hirk2022corporate} and \cite{vana2024multivariate}.

In a recent work \cite{jahn2025modeling}  proposed several regression-type models based on GARCH-type models for discrete valued time series. They extended their work by assuming a Gaussian copula to define the joint distribution of the multivariate ordinal time series and capturing the cross-dependence, while for large number of dimensions, the joint probability mass function is approximated by the mid-point approximation proposed by \cite{kazianka2010copula}.

There are two main reasons for the limited development of such models. First, there is not  an  easily implementable multivariate distribution for ordinal variables.  Second, an adequate multivariate framework must account not only for the autocorrelation within each time series, but also for the cross-correlation between different series.

Our approach builds on the model used in \cite{jahn2025modeling}  where the joint behaviour of many series using  copulas is considered. Copulas provide a convenient and powerful framework for constructing joint distributions for variables of arbitrary type and allow for a wide range of dependence structures. In our setting, the copula captures the contemporaneous (lag-0) cross-correlation between the series. Cross-correlations at higher lags, as well as serial dependence within each series, are incorporated by including past values of all time series in the marginal models. For the marginal distributions, any of the modelling approaches discussed above may be employed.
The main drawback of this approach is the need to evaluate a multivariate copula which can be painful and approximation may not resolve the problem.

Our proposal tries to circumvent this problem 
 through a pairwise likelihood approach, following the framework of \cite{fieuws2006pairwise}. Rather than fitting a single high-dimensional copula, we propose modelling all relevant bivariate relationships and approximating the full likelihood by a composite (pairwise) likelihood. This strategy substantially reduces computational complexity and offers greater flexibility, as it allows different copula families to be selected for different pairs of time series according to their specific dependence characteristics.

The paper is organized as follows: In \hyperref[sec2]{Section 2} we present how we can define the bivariate distribution of ordinal time series, how can we extend it in higher dimensions using pairwise likelihood and how we can estimate the multivariate model in a more efficient way. In \hyperref[sec3]{Section 3} we examine model's performance under different sample sizes. In \hyperref[sec4]{Section 4} we propose an approach for forecasting based on the multivariate model. The proposed methodology is applied to jointly model the unemployment level of six different countries (\hyperref[sec5]{Section 5}). In the last \hyperref[sec6]{Section 6}, we summarize the basic points of the methodology and possible extensions and improvements.

\section{Methodology}
\label{sec2}

\subsection{Multivariate model}

Assume that we have $K$ ordinal time series $Z_{1t},Z_{2t},\ldots,Z_{Kt}$, $t=1,\ldots,T$. Each ordinal time series takes values in $\mathcal{S}_k=\{s_{1},s_{2},\ldots,s_{d_k}\}$ where $s_{1}<s_{2}<\ldots<s_{d_k}$, $k=1,\ldots,K$. We assume that marginally each time series  follow the ordinal autoregressive logit model of order $p$ \citep{fokianos2003regression}, including lagged values of order $p$ from all time series of the system to capture auto-correlation and cross-correlation: 

\begin{eqnarray}
\label{Equation: Marginal model multivariate}
Z_{kt}|\mathcal{F}_{t-1} &\sim& Multinomial(1;\pi^{(k)}_{s_1t},\pi^{(k)}_{s_2t},\ldots,\pi^{(k)}_{s_{d_k}t})  \nonumber \\
  \logit(\gamma^{(k)}_{s_jt})&=&\alpha_{k0j}+\sum_{m=1}^{K} \bm{\alpha}_{km1}'\bm{\tilde{Z}}_{m,t-1}+\sum_{m=1}^{K}\bm{\alpha}_{km2}'\bm{\tilde{Z}}_{m,t-2}
+\ldots+\sum_{m=1}^{K}\bm{\alpha}_{kmp}'\bm{\tilde{Z}}_{m,t-p} 
\end{eqnarray}
where, $\mathcal{F}_{t-1}=\bm{\sigma}\{{Z_{1i},\ldots,Z_{Ki},i\leq t-1}\}$ is a $\sigma-$algebra containing all the history of the system of time series,$\gamma^{(k)}_{s_jt}=P(Z_{kt}\leq s_j|\mathcal{F}_{t-1})=\sum_{h=1}^{j}\pi^{(k)}_{s_ht}$, $\tilde{\bf{Z}}_{kt}$ the binary representation of $Z_{kt}$ (e.g. if $\mathcal{S}_1=\{1,2,3\}$ and $Z_{1t}=2$ then $\tilde{\bf{Z}}_{1t}=(0,1,0)^{'}$) 
for $j=1,\ldots,d_k-1$, $k=1,\ldots,K$, $t=2,\ldots,T$. $\bm{\alpha}'_{kmg}$ is a $(1 \times  (d_{m} - 1))$ is the vector of coefficients of each level of $Z_{m,t-g}$, $m=1,\ldots,K$, $g=1,\ldots,p$ for the model of $Z_{kt}$. Then one can use a multivariate copula to introduce the dependence between the $K$ series \citep[see][]{jahn2025modeling}. While this is feasible, the calculation of the joint probabilities implies $2^K$ terms leading to sever problems when $K$ increases. Here, while we assume that there is a multivariate copula that can generate the structure over the $K$ series, we will work with only the bivariate marginal models, namely we consider for all pairs the bivariate marginal distribution. 

 For a pair of variables $(Z_{rt},Z_{st})$, $r,s=1,\ldots,K \quad r\neq s$  their joint cumulative distribution function is given by:
$$F_{Z_{rt},Z_{st}}(z_{rt},z_{st}|\mathcal{F}_{t-1})=C(F_{Z_{rt}}(z_{rt}),F_{Z_{st}}(z_{st})|\mathcal{F}_{t-1},\phi)$$ where $F_{Z_{kt}}(\cdot)$ denotes the cumulative distribution function of $Z_{kt}|\mathcal{F}_{t-1}$, while  their joint probability mass function is given by differences, namely:
\begin{equation}
   \begin{split}
       f_{Z_{rt},Z_{st}}(z_{rt},z_{st}|{\cal{F}}_{t-1}) &=C(F_{Z_{rt}}(z_{rt}|\mathcal{F}_{t-1}),F_{Z_{st}}(z_{st}|{\cal{F}}_{t-1}),\phi)\\ &-C(F_{Z_{rt}}(z_{rt}-1|\mathcal{F}_{t-1}),F_{Z_{st}}(z_{st}|{\cal{F}}_{t-1}),\phi)\\
       &-C(F_{Z_{rt}}(z_{rt}|\mathcal{F}_{t-1}),F_{Z_{st}}(z_{st}-1|{\cal{F}}_{t-1}),\phi) \\ 
       &+C(F_{Z_{rt}}(z_{rt}-1|\mathcal{F}_{t-1}),F_{Z_{st}}(z_{st}-1|{\cal{F}}_{t-1}),\phi).
   \end{split} 
\end{equation}
Here $C(\cdot,\cdot,\phi)$ is any copula, $\phi$ denotes the copulas parameter(s).

We will base our estimation method on a pairwise likelihood, namely an objective function that uses using only the bivariate marginal distributions. 

 Pairwise likelihood \citep{cox2004note} approximates full likelihood by multiplying all possible bivariate models. So, for $K$-dimensional vector we need to specify $K(K-1)/2$ bivariate models.  In our case this is very helpful since we avoid the full specification of the multivariate model and we need only the bivariate models which are easier to handle. 

Then the conditional pairwise log-likelihood is defined as:
\begin{equation}
\ell_{prw}(\bm{\theta};\bm{z})=\sum_{t=2}^{T}\sum_{r=1}^{K-1}\sum_{s=r+1}^{K} \log f_{Z_{rt},Z_{st}}(z_{rt},z_{st};\bm{\theta}_{rs}|F_{t-1}) 
 = \sum_{t=2}^{T}\sum_{r=1}^{K-1}\sum_{s=r+1}^{K} \ell_{t}^{r,s}(\bm{\theta}_{rs})
    \label{Eq: Pairwise log-likelihood}
\end{equation}
Usage of pairwise likelihood not only reduces computational complexity of the model, but also allows for a more flexible modelling as we can use different copula family in each pair of time series according to the dependence structure that data implies. 

At this point it is important to highlight that in multivariate case,  marginal models will incorporate lagged values of all time series including in the system, not only of the series that participate in the bivariate model.
Also, we emphasize that our approach can be generalized with any other  model for ordinal time series and any copula to specify the cross-dependence structure.

\subsection{Estimation of multivariate model}
Optimizing conditional pairwise log-likelihood (\ref{Eq: Pairwise log-likelihood}) is hard. As an alternative approach, \cite{fieuws2006pairwise} showed that maximizing each bivariate model individually and combining the derived estimates are equivalent (see Appendix for some simulation results that support the  validity of the statement in our case) of maximizing the pairwise likelihood. Based on this methodology, estimation procedure consists of two steps. At the first step, we have to fit all $K(K-1)/2$ models by maximizing separately the terms of (\ref{Eq: Pairwise log-likelihood}) namely by maximizing:
\begin{equation}
     \ell^{r,s}({\bm{\theta_{rs}}})= \sum_{t=2}^{T}\log(f(z_{rt},z_{st};\bm{\theta_{rs}}|{\cal{F}}_{t-1}))=\sum_{t=2}^{T}\ell_t^{r,s}(\bm{\theta_{rs}})\
     \label{Eq: Bivariate loglik}
\end{equation}
for all pairs $r,s, r\neq s$. Note that this implies that some parameters may be estimated more than once. 
Then, at the second step, we have to average the individual estimates to obtain unique estimates for each parameter of the full joint model. The simplest way to combine them is by taking their mean value. However, with the purpose of improving properties of the final estimates we prefer using a weighted mean. Especially, we suggest the methodology presented in the work of \cite{hui2018sparse}. Assuming that $\tilde{\bm{\theta}}_{rs}$ is the estimator for the $(r,s)$ pair of variables, then we approximate each bivariate log-likelihood  in(\ref{Eq: Bivariate loglik}) using Taylor expansion of second order around $\tilde{\bm{\theta}}_{rs}$: 
\begin{equation}
\ell^{r,s}({\bm{\theta}_{rs}})\approx \ell^{r,s}(\tilde{\bm{\theta}}_{rs})+\frac{1}{2}(\tilde{\bm{\theta}}_{rs}-{\bm{\theta}_{rs}})^{T}H(\tilde{\bm{\theta}}_{rs})(\tilde{\bm{\theta}}_{rs}-{\bm{\theta}_{rs}}).
    \label{Eq: Taylor expansion}
\end{equation}
Thus, conditional pairwise log-likelihood in (\ref{Eq: Pairwise log-likelihood}) is approximated by: 
\begin{equation}
\ell_{aprw}(\bm{\theta})=-\frac{1}{2}\sum_{r=1}^{K-1}\sum_{r+1}^{K}(\bm{\theta}_{rs}-\tilde{\bm{\theta}}_{rs})^{T}H(\tilde{\bm{\theta}}_{rs})(\bm{\theta}_{rs}-\tilde{\bm{\theta}_{rs}}),
\label{Eq: approximate pairwise}
\end{equation}
where  $H(\tilde{\bm{\theta}}_{rs})=-\partial^{2}{\ell(\bm{\theta}_{rs})}/\partial{\bm{\theta}_{rs}\bm{\theta}_{rs}^{\tau}}$ is the negative Hessian matrix. 

Maximizing (\ref{Eq: approximate pairwise}), we  obtain a weighted mean of $\tilde{\bm{\theta}}_{rs}$. Augmenting properly the vector $\tilde{\bm{\theta}}_{rs}$
with zeroes, in the positions which do not associate with $(r,s)$ pair, we define $\tilde{\bm{\theta}}^{f}_{rs}$ the vector that has the same length as $\bm{\theta}$. We treat similarly Hessian matrix by adding zero rows and columns in positions that do not correspond to the pair $(r,s)$. The new matrix $H(\tilde{\bm{\theta}}^{f}_{rs})$ has
dimension dim($\bm{\theta}$)$\times$ dim($\bm{\theta}$). Then (\ref{Eq: approximate pairwise}) is written as: 

\begin{equation}
\ell_{aprw}(\bm{\theta})=-\frac{1}{2}\sum_{r=1}^{K-1}\sum_{r+1}^{K}(\bm{\theta}-\tilde{\bm{\theta}}^{f}_{rs})^{T}H(\tilde{\bm{\theta}}^{f}_{rs})(\bm{\theta}-\tilde{\bm{\theta}}^{f}_{rs})
\label{Eq: augmented approximate pairwise }
\end{equation}
Maximizing the above weighted least squares criterion (\ref{Eq: augmented approximate pairwise }) we obtain the weighted mean:

\begin{equation}
    \tilde{\bm{\theta}}_{wm}=\left \{\sum_{r=1}^{K-1}\sum_{s=r1}^{K} H(\tilde{\bm{\theta}}^{f}_{rs}) \right \}^{-1} \sum_{r=1}^{K-1}\sum_{s=r+1}^{K}H(\tilde{\bm{\theta}}^{f}_{rs}) \tilde{\bm{\theta}}^{f}_{rs}.
    \label{Eq: Weighted mean}
\end{equation}

\subsection{Standard error}
Pairwise likelihood belongs in the family of composite likelihoods, so we can benefit from the asymptotic distribution of the maximum composite likelihood estimator $\hat{\bm{\theta}}$ to calculate the standard errors \citep{fieuws2006pairwise}. More specifically, the asymptotic multivariate normal distribution of $\hat{\bm{\theta}}$, the vector that includes the estimates for the parameters of all bivariate models, is given by:

 \begin{align*}
     \sqrt{(T-k)}(\hat{\bm{\theta}}-\bm{\theta})\sim MVN(\bm{0},J^{-1}KJ^{-1})
 \end{align*}
 where $J$ is a block-diagonal matrix with diagonal blocks $J_{pp}$, and where $K$ is a symmetric matrix containing blocks $K_{pq}$, given by

\begin{align*}
 J_{pp}&=\frac{1}{T-k} \sum_{i=1}^{T-k}E \left( \frac{\partial^{2}\ell_{p_i}}{\partial \bm{\theta}_p \partial \bm{\theta}^{'}_p} \right )\\
 K_{pq}&=\frac{1}{T-k}\sum_{i=1}^{T-k}E \left (\frac{\partial\ell_{p_i}}{\partial \bm{\partial}_p}\frac{\partial \ell_{q_i}}{\partial \bm{\theta}^{'}_q} \right ), \quad p,q=1,\ldots,P
\end{align*}
where $P$ denotes the number of all possible bivariate models.  Since expected values are not known, we can replace them we the observed quantities by dropping expectation and replacing the estimated parameters. 

 In this way, we can estimate the standard error for each parameter in each bivariate model. However, our final estimates is the combination of the individual parameters, whether just a simple averaging or a weighted averaging according to (\ref{Eq: Weighted mean}). Assuming that $\bm{A}$ is the matrix with coefficients for weighting, it holds that $\hat{\tilde{\bm{\theta}}}_{wm}=\bm{A}\hat{\bm{\theta}}$ and then the asymptotic distribution of $\hat{\tilde{\bm{\theta}}}$ is given by: 

 \begin{align*}
 \sqrt{(T-k)}(\bm{A}\hat{\bm{\theta}}-\bm{A}\bm{\theta}) \sim MNV(\bm{0},\bm{A}\Sigma(\bm{\theta})\bm{A}^{'})  
 \end{align*}
where $\Sigma(\bm{\theta})=J^{-1}KJ^{-1}$
 For the simple averaging, matrix $\bm{A}$ gives the same weight in all individual estimates. For the weighted approach, weights are proportional to the correspond diagonal elements of Hessian matrices of the bivariate models.

\section{Simulation Study}
\label{sec3}
In this section we would like to study the performance of our method under different sample sizes. We assume a trivariate case. All models has three possible states $\mathcal{S}=\{1,2,3\}$, where $1<2<3$. Each model follows the ordinal autoregressive logit model of order $1$:

\begin{align*}
Z_{1t}|\mathcal{F}_{t-1} &\sim Multinomial(1;\pi^{(1)}_{1t},\pi^{(1)}_{2t},\pi^{(1)}_{3t}),\quad t=1,\ldots,T\\
\logit(\gamma^{(1)}_{1t})=\alpha^{(1)}_{01}&+\alpha_{11}I(Z_{1,t-1}=2)+\alpha_{12}I(Z_{1,t-1}=3)+\alpha_{13}I(Z_{2,t-1}=2)\\
&+\alpha_{14}I(Z_{2,t-1}=3)+\alpha_{15}I(Z_{3,t-1}=2)+\alpha_{16}I(Z_{3,t-1}=3)\\
\logit(\gamma^{(1)}_{2t})=\alpha^{(1)}_{02}&+\alpha_{11}I(Z_{1,t-1}=2)+\alpha_{12}I(Z_{1,t-1}=3)+\alpha_{13}I(Z_{2,t-1}=2)\\
&+\alpha_{14}I(Z_{2,t-1}=3)+\alpha_{15}I(Z_{3,t-1}=2)+\alpha_{16}I(Z_{3,t-1}=3)
\end{align*}
where, $\alpha^{(1)}_{01}=-0.50,\alpha^{(1)}_{02}=0.50,\alpha_{11}=0.50,\alpha_{12}=0.40,\alpha_{13}=0.15,\alpha_{14}=0.25,\alpha_{15}=0.10,\alpha_{16}=0.20$.

\begin{align*}
Z_{2t}|\mathcal{F}_{t-1} &\sim Multinomial(1;\pi^{(2)}_{1t},\pi^{(2)}_{2t},\pi^{(2)}_{3t})\\
\logit(\gamma^{(2)}_{1t})=\alpha^{(2)}_{01}&+\alpha_{21}I(Z_{1,t-1}=2)+\alpha_{22}I(Z_{1,t-1}=3)+\alpha_{22}I(Z_{2,t-1}=2)\\
&+\alpha_{24}I(Z_{2,t-1}=3)+\alpha_{25}I(Z_{3,t-1}=2)+\alpha_{26}I(Z_{3,t-1}=3)\\
\logit(\gamma^{(2)}_{2t})=\alpha^{(2)}_{02}&+\alpha_{21}I(Z_{1,t-1}=2)+\alpha_{22}I(Z_{1,t-1}=3)+\alpha_{23}I(Z_{2,t-1}=2)\\
&+\alpha_{24}I(Z_{2,t-1}=3)+\alpha_{25}I(Z_{3,t-1}=2)+\alpha_{26}I(Z_{3,t-1}=3)
\end{align*}
where, $\alpha^{(2)}_{01}=-0.30,\alpha^{(2)}_{02}=0.70,\alpha_{21}=0.15,\alpha_{22}=0.25,\alpha_{23}=0.30,\alpha_{24}=0.60,\alpha_{25}=0.25,\alpha_{26}=0.40$.

\begin{align*}
Z_{3t}|\mathcal{F}_{t-1} &\sim Multinomial(1;\pi^{(3)}_{1t},\pi^{(3)}_{2t},\pi^{(3)}_{3t})\\
\logit(\gamma^{(3)}_{1t})=\alpha^{(3)}_{01}&+\alpha_{31}I(Z_{1,t-1}=2)+\alpha_{32}I(Z_{1,t-1}=3)+\alpha_{33}I(Z_{2,t-1}=2)\\
&+\alpha_{34}I(Z_{2,t-1}=3)+\alpha_{35}I(Z_{3,t-1}=2)+\alpha_{36}I(Z_{3,t-1}=3)\\
\logit(\gamma^{(3)}_{2t})=\alpha^{(3)}_{02}&+\alpha_{31}I(Z_{1,t-1}=2)+\alpha_{32}I(Z_{1,t-1}=3)+\alpha_{33}I(Z_{2,t-1}=2)\\
&+\alpha_{34}I(Z_{2,t-1}=3)+\alpha_{35}I(Z_{3,t-1}=2)+\alpha_{36}I(Z_{3,t-1}=3)
\end{align*}
where, $\alpha^{(3)}_{01}=-0.40,\alpha^{(3)}_{02}=0.80,\alpha_{31}=0.20,\alpha_{32}=0.30,\alpha_{33}=0.15,\alpha_{34}=0.25,\alpha_{35}=0.40,\alpha_{36}=0.70$.\vspace{0.3cm}

In each case, $\mathcal{F}_{t-1}=\bm{\sigma}\{{Z_{1i},Z_{2i},Z_{3i},i\leq t-1}\}$ is a $\sigma-$algebra containing all the history of the three time series, $\gamma^{(k)}_{jt}=P(Z_{kt}\leq j|\mathcal{F}_{t-1})$ for $k=1,2,3$ and $j=1,2$ and $I(\cdot)$ is an indicator function. The joint distribution of $\bm{Z}_t=(Z_{1t},Z_{2t},Z_{3t})^{'}$ is defined by a trivariate Gumbel copula  with dependence parameter $\phi=2$. Thus, we assume that dependence between all possible pairs of time series is described by a common $\phi$. The trivariate Gumbel copula is defined as:
\begin{equation}
\label{Eq: Gumbel}
    C(u_1,u_2,u_3;\phi)=\exp \big[ -\big(-\log(u_1)^{\phi}-\log(u_2)^{\phi}-\log(u_3)^{\phi}\big)^{\frac{1}{\phi}}\big],  \quad \phi \in [1,\infty)
\end{equation}
We have simulated $100$ samples of sample size $T=100,500,1000$, with the purpose of examining and comparing the behavior of the two presented estimators: mean and weighted mean, under different conditions. The simulation procedure was based on the following algorithm:

\begin{algorithm}
\caption{Simulation from a trivariate model}
\begin{enumerate}
        \item Simulate $T+r$ values $(\bm{u}_t=(u_{1t},u_{2t},u_{3t})', ~ t=1,\ldots,T)$ from a trivariate Gumbel copula with $\phi=2$. The first $r$ observations are used as "burn-in" period. 
        \item Simulate marginal models as follows:\\ 
        $z_{1t}=F_{Z_{1t}}^{-1}(u_{1t})$ \quad for $i=1,\ldots,T$\\
        $z_{2t}=F_{Z_{2t}}^{-1}(u_{2t})$ \quad for $i=1,\ldots,T$\\
        $z_{3t}=F_{Z_{3t}}^{-1}(u_{3t})$ \quad for $i=1,\ldots,T$\\
        where $F_{Z_{kt}}(\cdot)$ is the conditional cumulative distribution function  ${Z_{kt}}|\mathcal{F}_{t-1}, ~ k=1,2,3$ 
        \item Remove $r$ first observations from each time series. The last $T$ observations is the final sample. 
\end{enumerate}
\end{algorithm}

\begin{figure}
\begin{center}
\includegraphics[scale=0.45]{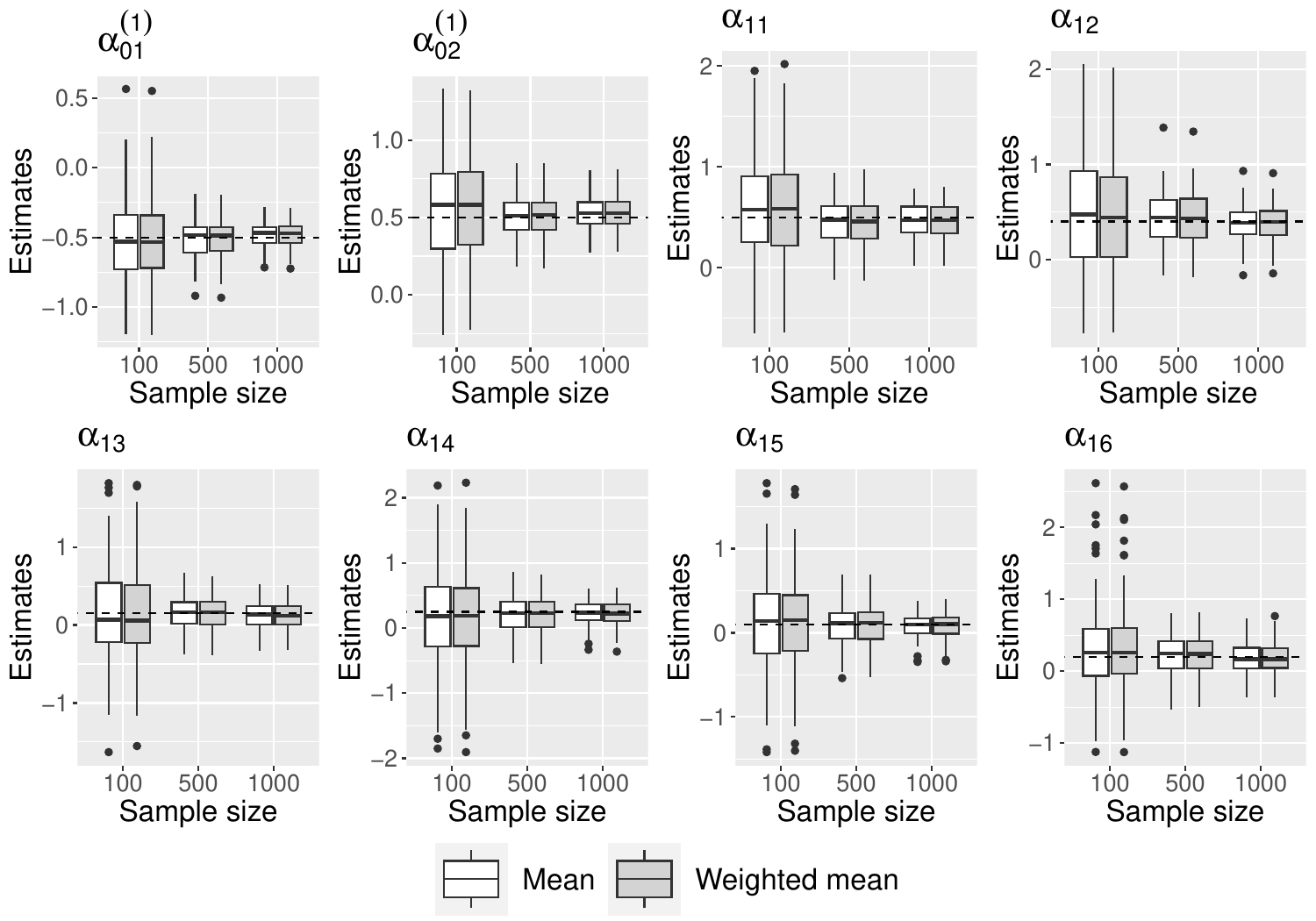}
\caption{\label{Figure: Z1 sim} Boxplots of estimates of $Z_{1t}$ model for different sample sizes. Dashed line is the true value of the parameter.}
\end{center}
\end{figure}

\begin{figure}
\begin{center}
\includegraphics[scale=0.45]{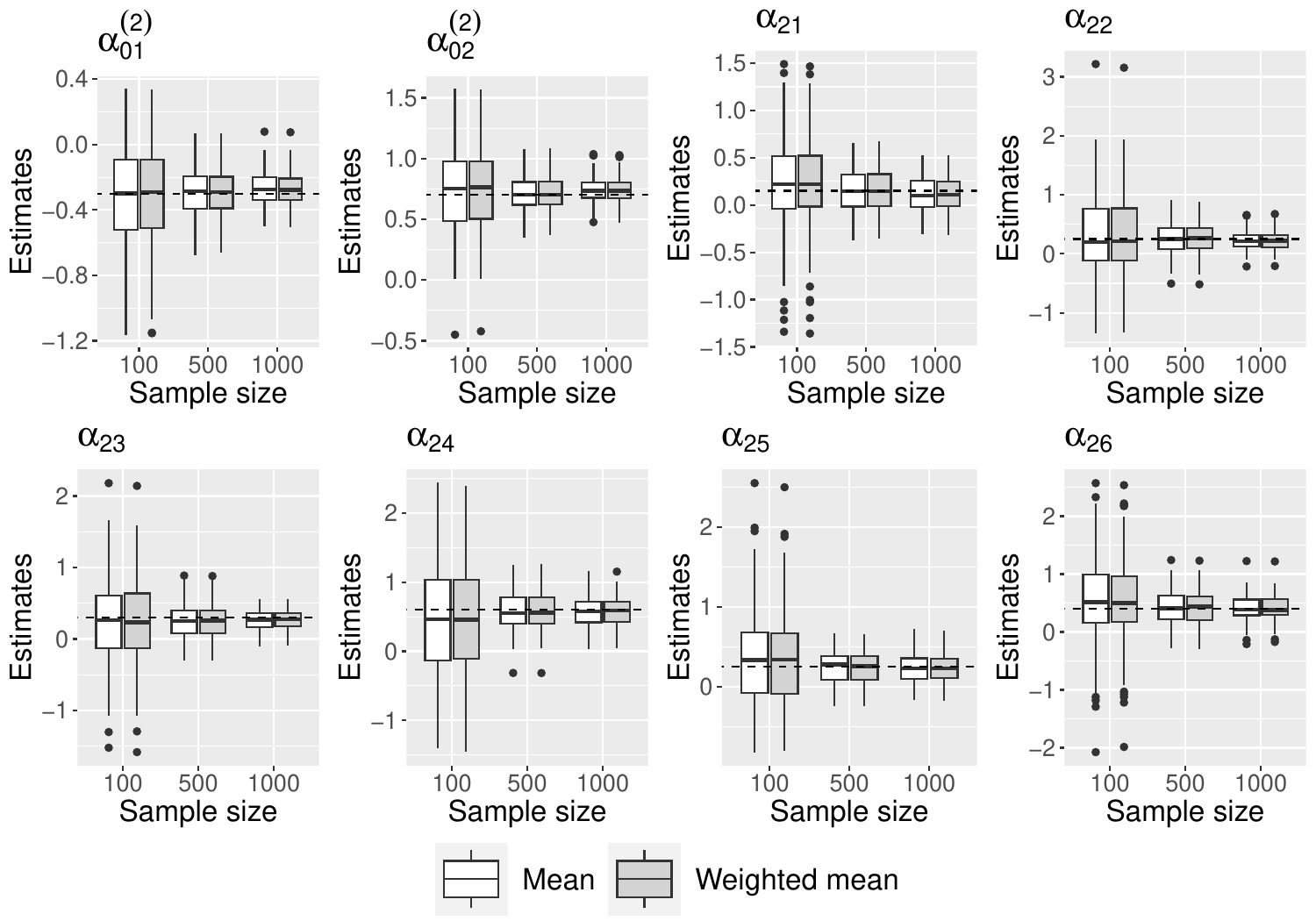}
\caption{\label{Figure: Z2 sim} Boxplots of estimates of $Z_{2t}$ model for different sample sizes. Dashed line is the true value of the parameter.}
\end{center}
\end{figure}

\begin{figure}
\begin{center}
\includegraphics[scale=0.45]{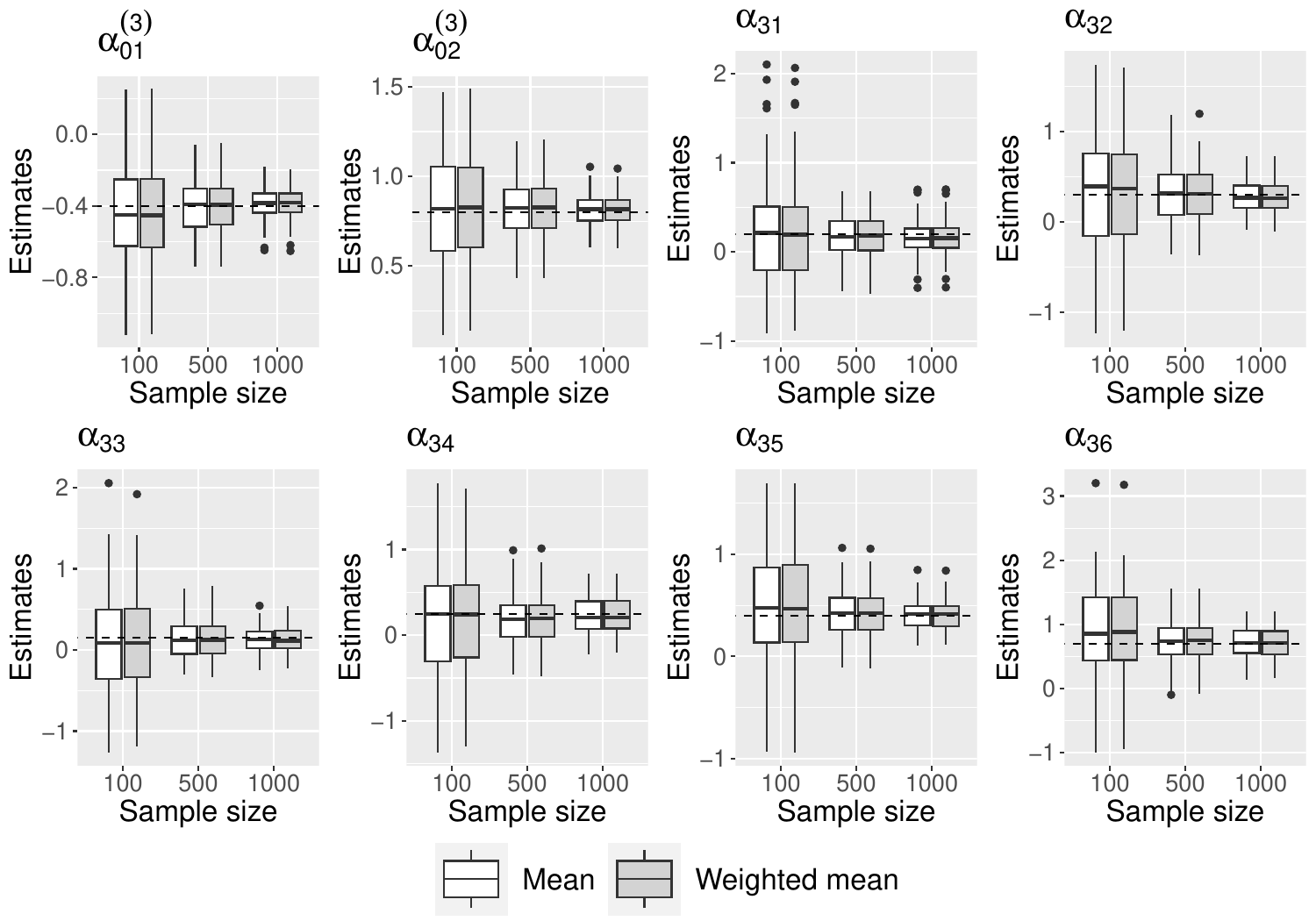}
\caption{\label{Figure:Z3 sim} Boxplots of estimates of $Z_{3t}$ model for different sample sizes. Dashed line is the true value of the parameter.}
\end{center}
\end{figure}

\begin{figure}
\begin{center}
\includegraphics[scale=0.45]{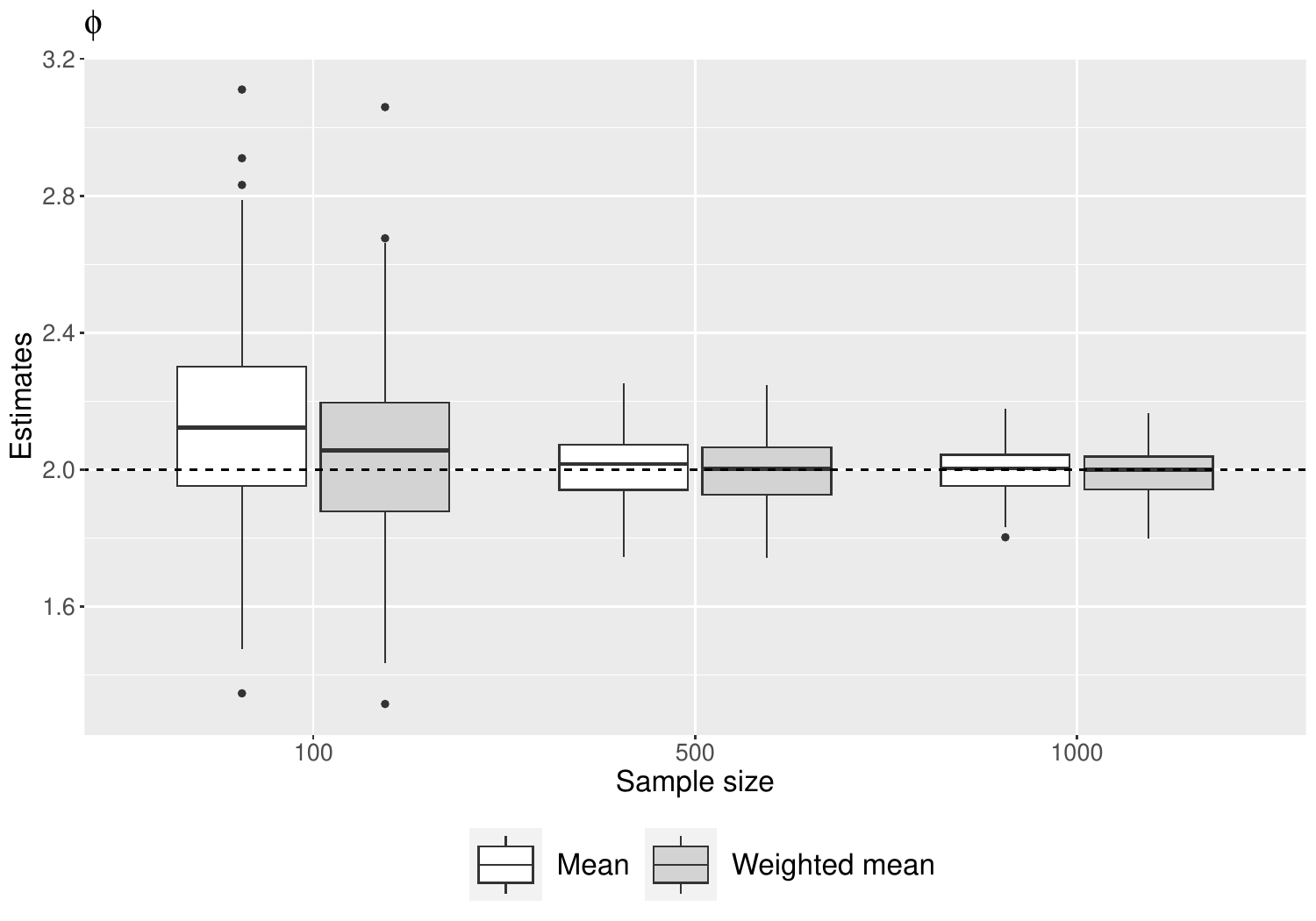}
\caption{\label{Figure:phi sim} Boxplots of estimates of $\phi$ for different sample sizes. Dashed line is the true value of the parameter.}
\end{center}
\end{figure}

\begin{figure}
\begin{center}
\includegraphics[scale=0.45]{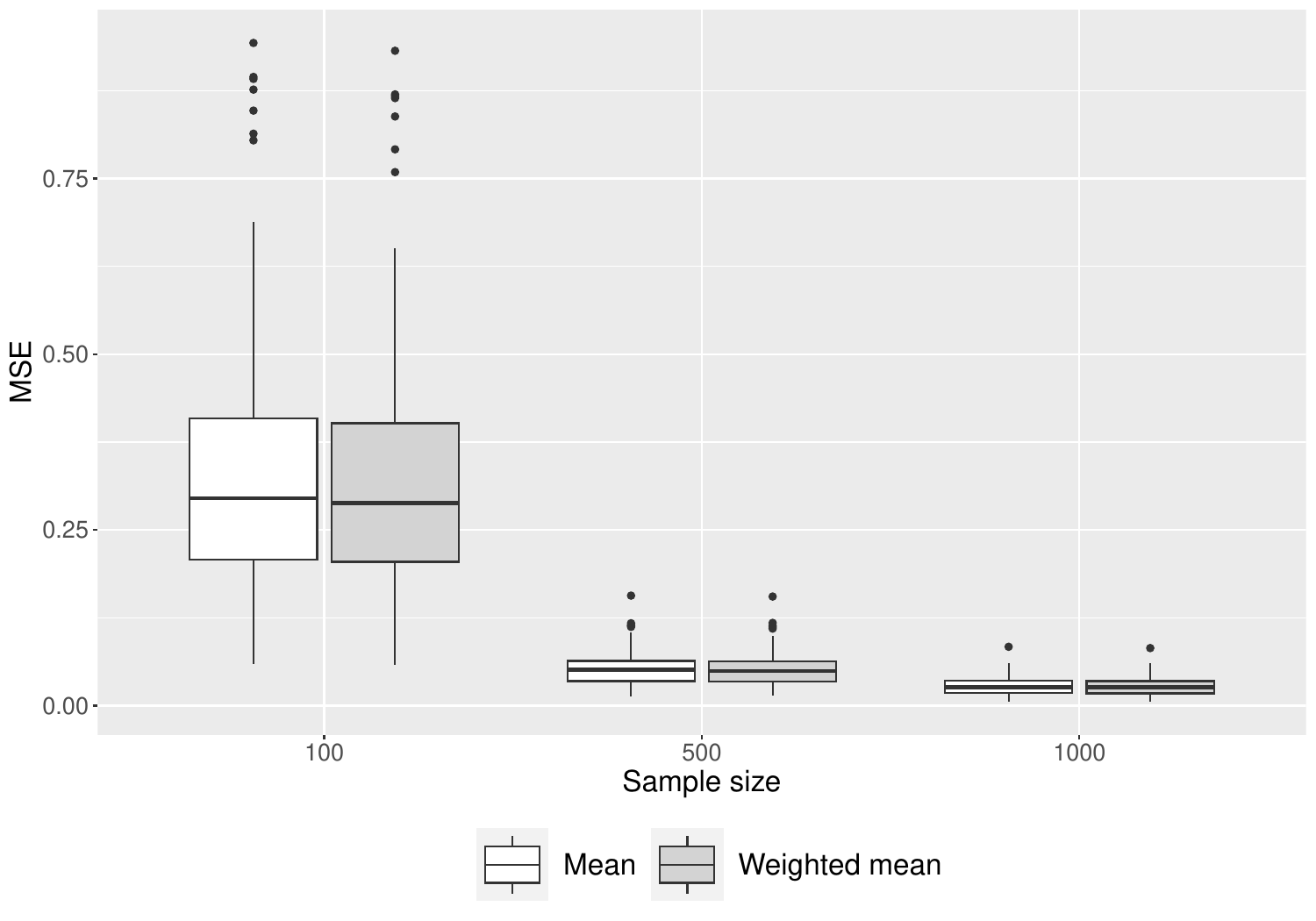}
\caption{\label{Figure: MSE} Boxplots of  MSE for mean and weighted mean for each sample size}
\end{center}
\end{figure}

 Figures \ref{Figure: Z1 sim}-\ref{Figure:Z3 sim} show the results for the regression coefficients, while  Figure \ref{Figure:phi sim} presents the results for the dependence parameters.  
 One can see that in all the cases, the 
 medians of boxplots are very close to the true values indicating unbiased estimators. In addition, in each case as sample size increases, variance decreases, implying consistent estimators. Comparing mean and weighted mean, there are very slight differences in most of the cases. However an exception is $\phi$ parameter in Figure \ref{Figure:phi sim}, where weighted mean outperform mean in each sample size. 
 We also provide the Mean Squared Error (MSE) for the whole vector of parameters per sample. Especially, for each iteration we have:  
\begin{align*}
    MSE=\frac{\sum_{j=1}^{25}(\theta_j-\hat{\theta}_j)^2}{25}
\end{align*}
 where, $\bm{\theta}=(\theta_1,\theta_2,\ldots,\theta_{25})^{'}$ the vector of true parameters and $\bm{\hat{\theta}}=(\hat{\theta}_1,\hat{\theta}_2,\ldots,\hat{\theta}_{25})^{'}$ the corresponding vector of parameter estimates. The true parameter vector in this example is
\begin{eqnarray*}
\bm{\theta} &=&(\alpha^{(1)}_{01},\alpha^{(1)}_{02},\alpha_{11},\alpha_{12},\alpha_{13},\alpha_{14},\alpha_{15},\alpha_{16}, \alpha^{(2)}_{01},\alpha^{(2)}_{02},\alpha_{21},\alpha_{22},\alpha_{23},\\ && \alpha_{24},\alpha_{25},\alpha_{26}, 
\alpha^{(3)}_{01},\alpha^{(3)}_{02},\alpha_{31},
\alpha_{32},\alpha_{33},\alpha_{34},\alpha_{35},\alpha_{36},\phi)^{'}
\end{eqnarray*}
and $\bm{\hat{\theta}}$ is defined analogously with each component replaced by its estimate.
 The results are presented in Figure \ref{Figure: MSE}, indicating that as sample size increases MSE goes to zero for both estimators, while they present similar results.

\section{Forecasting}
\label{sec4}
Forecasting is strongly related to any time series model. A typical  the motivation for developing  time series models is to make predictions based on the current available data. However, usage of pairwise likelihood approach makes forecasting procedure to be ambiguous due to the limited/conflicting  available information. More specifically, in our case, the multivariate distribution is not available and we try to approximate it by the product of all possible bivariate models. Thus, we can predict each time series based on bivariate models. This means that we have more than one forecast for each time series. 
So, the problem is how we can combine the individual forecasts to obtain a unique one. The simplest thing we can do is just averaging them. However, the problem of combining forecasts of different models has occupied researchers and a great variety of methods have been proposed. \cite{wang2023forecast} provide an extensive review of the methods. 

Since, we model ordinal time series, another topic we have to consider is whether we would like to forecast the $\pi_{jt}$'s for each state or the state itself, for a number of time points ahead. In each case the combination approach should be different. For example, if the focus is on the probabilities, then we can use the mean value of the individual probabilities, from bivariate models, as a unique forecast. On the other hand, if the case is to predict the following states, then we can handle it in two ways. Firstly, based on the combined  probabilities $\pi^{*}_{jt}$, we can take as state prediction the mode or the median of the distribution, since we work with ordinal time series. Median is the smallest state where the cumulative probability is equal or greater than 0.5, while mode is the state with highest $\hat \pi_{jt}$. Alternatively, we can use mode or median to achieve a state prediction from each bivariate case and then find the mode of them as the final predicted state. More details about forecasting can be found on the Application (\hyperlink{sec5}{Section 5}).

\section{Application: Unemployment Level}
\label{sec5}

\subsection{About Data}
\begin{figure}
\begin{center}
\includegraphics[scale=0.42]{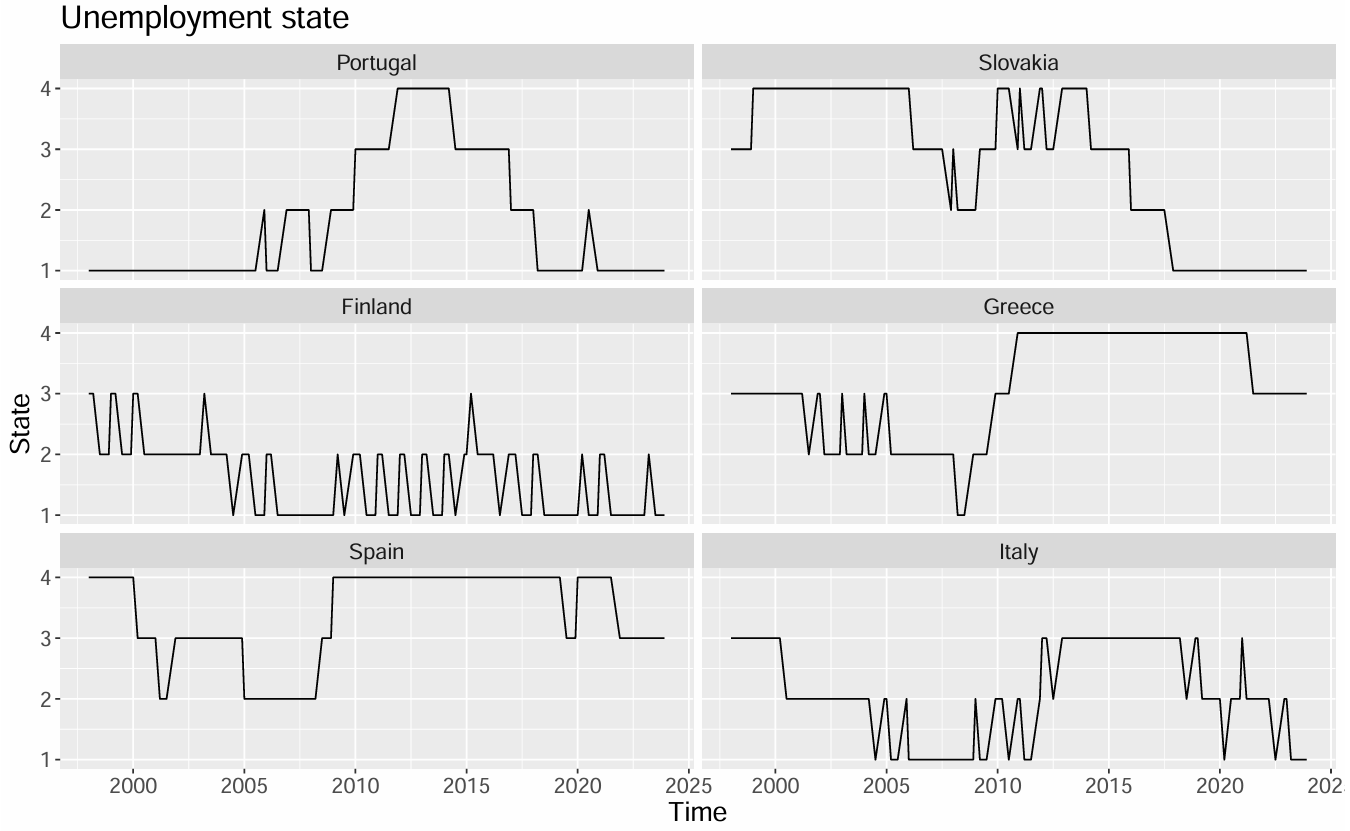}
\caption{\label{Fig. Data} Unemployment state per quarter for six countries from 1998 to 2023}
\end{center}
\end{figure}

The data refer to the unemployment rate for each quarter from 1998-2023 for six countries: Portugal, Slovakia, Finland, Greece, Spain and Italy, denoted by $Y_{kt}$ for $k=1,\ldots,6$ and $t=1,\ldots,104$. The data are provided by Eurostat. We assume that unemployment rate of each country is affected not only by previous values of itself but also from current and previous values of the rest countries. For simplicity we consider only values at lag $1$. However, with the purpose of making the values comparable, we propose discretize them based on the quantiles of the whole set of unemployment rates. Then, we convert continuous time series $(Y_{kt})$ into ordinal time series $(Z_{kt})$ with $4$ possible states, 
$Z_{kt}\in \mathcal{S}=\{1,2,3,4\}, \quad k=1,\ldots,6$, where $1<2<3<4$ based on the following rule:

\begin{equation*} 
 Z_{kt}= \left\{
\begin{array}{ll}
      1, & 3.8\leq Y_{kt} \leq 7.88 \\
      2, & 7.88 < Y_{kt} \leq 10.4\\
      3, & 10.4 < Y_{kt} \leq 13.9 \\
      4, & 13.9 < Y_{kt} \leq 27.9 \\
\end{array} 
\right. 
\end{equation*}
Discretizing into broad intervals reduces spurious precision and improves comparability. The resulting ordinal variable captures economically meaningful shifts in labor-market conditions while remaining robust to transitory fluctuations and crisis-driven outliers.
The final data are presented in Figure \ref{Fig. Data}, where we can notice that Spain never takes the state "1", while Finland and Italy  never exceeds state "3". This is something that we have to take into consideration during modelling, since it affects the marginal model of each series.   

To capture the cross-correlation between the countries we need to fit a multivariate model, like that we have proposed. Nevertheless, before fitting the model we will examine whether there are correlations, to ensure that the usage of a multivariate model is plausible. 
We selected to use Frank copula to capture both positive and negative dependence. We use Kendall's $\tau$ to quantify autocorrelation and cross-correlation, which is a common measure in case of ordinal data. Other measures for serial dependence and cross-correlation in case of ordinal time series can be found in \cite{weiss2019distance}. According to Table \ref{Tab. Cor lag 1 Multivariate} all countries except for Finland present very high autocorrelation at lag $1$. As correlation between countries at lag $0$ is concerned (Table \ref{Tab. Cor lag 0 Multivariare}), for the pairs Spain-Greece, Greece-Italy and Spain-Italy we have the strongest associations. On the other hand, there are pairs where the correlation is close to zero like Finland-Portugal, Greece-Finland and Italy-Slovakia.  The existence of strong temporal and cross correlation makes the used of multivariate time series models plausible. 

\begin{table}
\centering
\begin{tabular}{lcccccc} 
 \hline
          & $Z_{1,t-1}$ & $Z_{2,t-1}$ & $Z_{3,t-1}$& $Z_{4,t-1}$ & $Z_{5,t-1}$ & $Z_{6,t-1}$ \\ [0.5ex] 
 \hline
 $Z_{1t}$-Portugal & 0.891 & 0.142 & -0.097 & 0.36 & 0.404 & 0.161 \\ 
 $Z_{2t}$-Slovakia & 0.101 & 0.886 & 0.365  & -0.26& -0.126& 0.026\\
 $Z_{3t}$-Finland  & -0.023& 0.413 & 0.470 & 0.029 & 0.189 & 0.425\\
 $Z_{4t}$-Greece   & 0.443 & -0.252 & -0.043 & 0.891 & 0.720 & 0.532\\
 $Z_{5t}$-Spain    & 0.452 & -0.172 & 0.111 & 0.648 & 0.902 & 0.561\\
 $Z_{6t}$-Italy    & 0.289 & 0.041 & 0.234 & 0.560 & 0.636 & 0.748\\
 \hline
\end{tabular}
\caption{\label{Tab. Cor lag 1 Multivariate}Correlation matrix of time series at lag $1$ based on Kendall's tau.}
\end{table} 

\begin{table}
\centering
\begin{tabular}{lcccccc} 
 \hline
          & $Z_{1t}$ & $Z_{2t}$ & $Z_{3t}$ & $Z_{4t}$ & $Z_{5t}$ & $Z_{6t}$  \\ [0.5ex] 
 \hline
 $Z_{1t}$-Portugal & 1 &&&&& \\ 
 $Z_{2t}$-Slovakia &0.128 &1 &&&& \\
 $Z_{3t}$-Finland  &-0.079 & 0.408 &1  &&&\\
 $Z_{4t}$-Greece   &0.405  & -0.258 & 0.013 &1 &&  \\ 
 $Z_{5t}$-Spain    &0.425 & -0.143 & 0.158 & 0.688 &1&  \\
 $Z_{6t}$-Italy    &0.249  & 0.050 & 0.380 & 0.567 & 0.607& 1  \\
 \hline
\end{tabular}
\caption{\label{Tab. Cor lag 0 Multivariare}Correlation matrix of time series based on Kendall's tau}
\end{table}

\subsection{Model specification and estimation}

Marginally for each time series we consider the ordinal autoregressive logit model of order $1$. We also consider the linear effect of time series to keep the number of parameters low. Increasing the number of countries to be considered in the multivariate model leads to essential increase of parameters. For each new country we have to define an extra marginal model, and we also have to incorporate lagged values of this country in the marginal models of all the time series we study. In addition, the number of bivariate models also increases, and hence, we need to estimate extra copula's parameters to account for the association.
Our model takes the form:

$$Z_{kt}|\mathcal{F}_{t-1} \sim Multinomial(1;\pi^{(k)}_{1t},\pi^{(k)}_{2t},\pi^{(k)}_{3t},\pi^{(k)}_{4t})$$
\begin{align*}
\logit(\gamma^{(k)}_{jt})=\alpha_{k0j}+\alpha_{k1}Z_{1,t-1}+\alpha_{k2}Z_{2,t-1}+\alpha_{k3}Z_{3,t-1}\\
+\alpha_{k4}Z_{4,t-1}+\alpha_{k5}Z_{5,t-1}+\alpha_{k6}Z_{6,t-1}
\end{align*}
for $k=1,\ldots,6,~$ $j=1,2,3,~$ $t=2,\ldots,104$. $\mathcal{F}_{t-1}=\bm{\sigma}\{{Z_{1i},\ldots,Z_{6i},i\leq t-1}\}$ is a $\sigma-$algebra containing all the history of the system and $\gamma^{(k)}_{jt}=P(Z_{kt}\leq j|\mathcal{F}_{t-1})$, for $k=1,\ldots,6$ and $j=1,2,3$

To define the bivariate models, we use Frank copula
\begin{align*}
C_F(u,v;\phi)=-\frac{1}{\phi}\log \left[ 1+ \frac{(\exp(-\phi u)-1)(\exp(-\phi v)-1)}{\exp(-\phi)-1} \right],\quad \phi \in \mathbb{R} \setminus \{0\}  
\end{align*}
to account for the existence of positive and negative correlations as shown in Table \ref{Tab. Cor lag 0 Multivariare}.  Fitting our proposed model, the estimated parameters of the marginal models are presented in Table \ref{Tab:MultMeanMerged}. Table \ref{Tab. phi Multivariate} includes copulas' parameters. We also record the standard error of the estimators based on the asymptotic distribution of the pseudo-likelihood estimators (\citep{varin2011overview, fieuws2006pairwise}). The statistical significance of the parameters can be examined based on Wald test \citep{varin2011overview}.

\begin{table}[H]
\centering
\begin{tabularx}{\textwidth}{c *{6}{>{\centering\arraybackslash}X}}
 \hline
          & $Z_{1t}$ & $Z_{2t}$ & $Z_{3t}$ & $Z_{4t}$ & $Z_{5t}$ & $Z_{6t}$ \\  
 \hline
 \multicolumn{7}{c}{Mean} \\
 \hline
 $\alpha_{01}$ & $\underset{(4.552)}{13.271}$ & $\underset{(2.312)}{7.738}$ & $\underset{(1.757)}{9.103}$ & $\underset{(2.181)}{9.19}$ & - & $\underset{(2.592)}{11.041}$ \\
 $\alpha_{02}$ & $\underset{(5.013)}{18.335}$ & $\underset{(2.714)}{11.263}$ & $\underset{(2.066)}{14.033}$ & $\underset{(2.922)}{15.599}$ & $\underset{(2.184)}{11.404}$ & $\underset{(2.995)}{15.190}$ \\
 $\alpha_{03}$ & $\underset{(6.573)}{25.358}$ & $\underset{(2.874)}{15.345}$ & - & $\underset{(3.588)}{21.688}$ & $\underset{(2.860)}{17.595}$ & - \\
 $Z_{1,t-1}$   & $\underset{(0.726)}{-5.248}$ & $\underset{(0.397)}{0.175}$ & $\underset{(0.373)}{1.215}$ & $\underset{(0.437)}{-0.692}$ & $\underset{(0.571)}{-2.006}$ & $\underset{(0.392)}{0.325}$ \\ 
 $Z_{2,t-1}$   & $\underset{(0.621)}{-1.041}$ & $\underset{(0.604)}{-4.343}$ & $\underset{(0.329)}{-1.485}$ & $\underset{(0.399)}{-0.056}$ & $\underset{(0.400)}{1.043}$ & $\underset{(0.328)}{-0.796}$ \\ 
 $Z_{3,t-1}$   & $\underset{(0.646)}{0.675}$ & $\underset{(0.572)}{0.314}$ & $\underset{(0.525)}{-0.028}$ & $\underset{(0.707)}{1.242}$ & $\underset{(0.780)}{-0.618}$ & $\underset{(0.552)}{0.858}$ \\ 
 $Z_{4,t-1}$   & $\underset{(1.229)}{-0.862}$ & $\underset{(0.619)}{0.733}$ & $\underset{(0.507)}{0.133}$ & $\underset{(0.800)}{-4.065}$ & $\underset{(1.012)}{1.085}$ & $\underset{(0.666)}{-0.783}$ \\ 
 $Z_{5,t-1}$   & $\underset{(0.764)}{-1.391}$ & $\underset{(0.674)}{-0.902}$ & $\underset{(0.492)}{-1.236}$ & $\underset{(0.716)}{-1.934}$ & $\underset{(1.311)}{-4.963}$ & $\underset{(0.582)}{-1.627}$ \\ 
 $Z_{6,t-1}$   & $\underset{(0.868)}{1.973}$ & $\underset{(0.760)}{-0.222}$ & $\underset{(0.500)}{-1.760}$ & $\underset{(0.515)}{-0.478}$ & $\underset{(0.467)}{-1.139}$ & $\underset{(0.532)}{-2.554}$ \\ 
 
 \hline
 \multicolumn{7}{c}{Weighted Mean} \\
 \hline
 $\alpha_{01}$ & $\underset{(4.544)}{13.264}$ & $\underset{(2.319)}{7.764}$ & $\underset{(1.756)}{9.104}$ & $\underset{(2.181)}{9.196}$ & - & $\underset{(2.592)}{11.041}$ \\
 $\alpha_{02}$ & $\underset{(5.014)}{18.339}$ & $\underset{(2.722)}{11.273}$ & $\underset{(2.084)}{14.026}$ & $\underset{(2.914)}{15.559}$ & $\underset{(2.190)}{11.404}$ & $\underset{(2.995)}{15.190}$ \\
 $\alpha_{03}$ & $\underset{(6.584)}{25.375}$ & $\underset{(2.899)}{15.412}$ & - & $\underset{(3.565)}{21.595}$ & $\underset{(2.884)}{17.611}$ & - \\
 $Z_{1,t-1}$   & $\underset{(0.726)}{-5.248}$ & $\underset{(0.396)}{0.176}$ & $\underset{(0.377)}{1.212}$ & $\underset{(0.435)}{-0.689}$ & $\underset{(0.586)}{-1.971}$ & $\underset{(0.392)}{0.325}$ \\ 
 $Z_{2,t-1}$   & $\underset{(0.620)}{-1.041}$ & $\underset{(0.604)}{-4.343}$ & $\underset{(0.329)}{-1.486}$ & $\underset{(0.404)}{-0.064}$ & $\underset{(0.400)}{1.043}$ & $\underset{(0.328)}{-0.796}$ \\ 
 $Z_{3,t-1}$   & $\underset{(0.646)}{0.676}$ & $\underset{(0.576)}{0.310}$ & $\underset{(0.525)}{-0.028}$ & $\underset{(0.708)}{1.243}$ & $\underset{(0.777)}{-0.628}$ & $\underset{(0.552)}{0.858}$ \\ 
 $Z_{4,t-1}$   & $\underset{(1.229)}{-0.863}$ & $\underset{(0.616)}{0.731}$ & $\underset{(0.512)}{0.128}$ & $\underset{(0.799)}{-4.066}$ & $\underset{(1.024)}{1.111}$ & $\underset{(0.666)}{-0.783}$ \\ 
 $Z_{5,t-1}$   & $\underset{(0.763)}{-1.392}$ & $\underset{(0.674)}{-0.901}$ & $\underset{(0.492)}{-1.241}$ & $\underset{(0.729)}{-1.938}$ & $\underset{(1.309)}{-4.951}$ & $\underset{(0.582)}{-1.627}$ \\ 
 $Z_{6,t-1}$   & $\underset{(0.868)}{1.973}$ & $\underset{(0.766)}{-0.226}$ & $\underset{(0.504)}{-1.758}$ & $\underset{(0.514)}{-0.516}$ & $\underset{(0.477)}{-1.409}$ & $\underset{(0.832)}{-2.554}$ \\ 
 \hline 
\end{tabularx}
\caption{\label{Tab:MultMeanMerged} Final estimates and standard errors for different methods of estimations' synthesis}
\end{table}

\begin{table}
\centering
\begin{tabular}{ccccccc} 
 \hline
          & $Z_{1t}$ & $Z_{2t}$ & $Z_{3t}$ & $Z_{4t}$ & $Z_{5t}$ & $Z_{6t}$  \\ [0.5ex] 
 \hline
 $Z_{2t}$ &$\underset{(1.722)}{0.867}$ & &&&& \\
 $Z_{3t}$&$\underset{(1.843)}{-2.733}$ & $\underset{(1.316)}{1.410}$ &  &&&\\
 $Z_{4t}$&$\underset{(2.215)}{0.124}$  & $\underset{(2.367)}{-1.661}$ & $\underset{(1.533)}{-0.875}$ & &&  \\ 
 $Z_{5t}$ &$\underset{(2.991)}{3.636}$ & $\underset{(2.015)}{-0.350}$ & $\underset{(1.730)}{0.943}$ & $\underset{(2.369)}{1.427}$ &&  \\
 $Z_{6t}$ &$\underset{(1.982)}{2.488}$  & $\underset{(1.869)}{0.144}$ & $\underset{(1.328)}{0.964}$ & $\underset{(2.015)}{4.410}$ & $\underset{(2.033)}{4.676}$&   \\
 \hline
\end{tabular}
\caption{\label{Tab. phi Multivariate} Estimates of copula’s parameter and standard errors}
\end{table}

\subsection{Forecasting}
Forecasting is also interesting using the model developed. 
For this reason, now we assume that we have observed the time series until 2020 ($t=1,\ldots,T=92$) and we would like to forecast the unemployment state for each quarter of the period 2021-2023. Based on the observed data until 2020, we estimate the model. Then, to forecast $\bm{Z}_{T+h}=(Z_{1,T+h},Z_{2,T+h},Z_{3,T+h},Z_{4,T+h},Z_{5,T+h},Z_{6,T+h})^{'}$, for $h=1,\ldots,12$, we need to find the marginal probabilities $\pi_{jt}^{(k)}$ for $k=1,\ldots,6$ from each bivariate model, conditional to the previous values of the system. For example, consider the time series $Z_{1t}$, we get
\begin{align*}
    \hat{P}^{(1,2)}(Z_{1,T+h}=i|\bm{Z}_{T+h-1})=\sum_{j=1}^{4} \hat P(Z_{1,T+h}=i,Z_{2,T+h=j}|\bm{Z}_{T+h-1}),\\
    \hat{P}^{(1,3)}(Z_{1,T+h}=i|\bm{Z}_{T+h-1})=\sum_{j=1}^{4}\hat{P}(Z_{1,T+h}=i,Z_{3,T+h=j}|\bm{Z}_{T+h-1})\\
    \hat{P}^{(1,4)}(Z_{1,T+h}=i|\bm{Z}_{T+h-1})=\sum_{j=1}^{4}\hat{P}(Z_{1,T+h}=i,Z_{4,T+h=j}|\bm{Z}_{T+h-1})\\
    \hat{P}^{(1,5)}(Z_{1,T+h}=i|\bm{Z}_{t+h-1})=\sum_{j=1}^{4}\hat{P}(Z_{1,T+h}=i,Z_{5,t+h=j}|\bm{Z}_{T+h-1})\\
    \hat{P}^{(1,6)}(Z_{1,T+h}=i|\bm{Z}_{T+h-1})=\sum_{j=1}^{4}\hat{P}(Z_{1,T+h}=i,Z_{6,T+h=j}|\bm{Z}_{T+h-1})
\end{align*}
for $i=1,\ldots,4$.
 The superscripts indicate the bivariate pair model used for the prediction, so $P^{(1,2)}$ implies the bivariate model with variables $Z_{1t}$ and $Z_{2t}$.

Then, a forecast $\hat{Z}_{1,T+h}$ can occur in two ways: whether we can take a random draw from each distribution of $Z_{1,T+h}|\bm{Z}_{T+h-1}$ from $(1,2)$,$(1,3)$,$(1,4)$,$(1,5)$ and $(1,6)$ and as final forecast keep the mode of them or in case of ties just choose randomly one of them (Method A). Alternatively, we can combine the individual probabilities by taking their mean, leading to combine probabilities:
\begin{small}
  \begin{align*}
     \hat{P}^{*}(Z_{1,T+h}=i|\bm{Z}_{T+h-1})
     =\frac{\sum_{s=2}^{6}\hat{P}^{(1,s)}(Z_{1,T+h}=i|\bm{Z}_{T+h-1})}{5}. 
\end{align*}  
\end{small}
Then, $\hat{Z}_{1,T+h}$ is a random draw from the above final distribution (Method B). We repeat the above procedures  for $B=10000$ times. As final forecast at  each step $h$ we consider the most frequent state of the $B=10000$ repetitions. We follow the same procedure for forecasting the rest $Z_{kt}$, for $k=2,\ldots,6$. The results for both methods are presented in Tables \ref{Tab. Forc_ProbsA}-\ref{Tab. Forc_StatesB}. 
Based on them, the two methods seem to provide similar results. Especially, for Portugal and Slovakia both of them perform very well providing the true state with very high probability, while for Finland we also have satisfactory results. For Greece, Spain and Italy the results are poor. However, as Greece and Spain are concerned based on Figure \ref{Fig. Data}, the two countries insist staying in the same state for long periods. For this type of data, there are more appropriate models in the literature like DAR model \citep{weiss2008measuring,nalpantidi2025bivariate}, that captures long runs of repeating values.

The proposed methodology is general enough to allow assuming different marginal models for the time series of the system depending on the form of the data. For Italy, probabilities $\hat{\pi}^{(6)}_{2t}$, $\hat{\pi}^{(6)}_{3t}$ become quite close from $h=3$ and then. Using the mode as final forecast in cases like that maybe is not the best option, so a further investigation is needed. Finally, in each case, we provide forecasts for the next three years. However, these projections may
not be entirely realistic, as they extend far into the future.

\begin{table}
\centering
\begin{tabular}{|c|cccc|cccc|ccc|} 
\hline 
  & & $Z_{1,t+h}$ & &  & & $Z_{2,t+h}$  & & & &   $Z_{3,t+h}$&  \\
 \hline
    $h$ & 1 & 2 & 3 & 4 & 1 & 2 & 3 & 4 & 1 &2 &3   \\           
 \hline
1& 0.993 & 0.007 & 0 & 0 & 0.993 & 0.007 & 0 & 0 &  0.911& 0.088 &0\\
2& 0.986 & 0.014 & 0 & 0 & 0.987 & 0.013 & 0 & 0 & 0.812 & 0.188 &0\\
3& 0.979 & 0.020 & 0 & 0 & 0.981 & 0.018 & 0.001 & 0 & 0.727 & 0.273 & 0\\
4& 0.975 & 0.023 & 0.001 &0 &0.976 & 0.022 & 0.002 & 0 & 0.670 & 0.329 & 0\\
5& 0.970 & 0.028 & 0.002 & 0&0.970 & 0.028 & 0.002 & 0 & 0.621 & 0.378 & 0\\
6& 0.965 & 0.031 & 0.003 & 0 &0.965 & 0.032 & 0.004 & 0&  0.594 &0.405 & 0\\
7& 0.962 & 0.034 & 0.004 & 0 & 0.960 & 0.035 & 0.005 & 0&  0.566& 0.433 & 0.001\\
8&0.958 & 0.036 & 0.005 & 0 & 0.955 & 0.038 & 0.006 & 0.001& 0.544 & 0.455 & 0.001\\
9& 0.958 & 0.036 & 0.006 & 0 & 0.952 & 0.039 & 0.007 & 0.001& 0.530 & 0.468 & 0.002\\
10& 0.956 & 0.037 & 0.007 & 0 & 0.948 & 0.042 & 0.009 & 0.001& 0.528 & 0.470 & 0.002\\
11&0.955 &0.038 & 0.007 & 0 & 0.943 & 0.045 & 0.011 & 0.002& 0.518 & 0.479 & 0.003\\ 
12&0.953 & 0.040 & 0.008 & 0 & 0.938 & 0.047 & 0.012 & 0.002& 0.498 & 0.500 &0.002\\
 \hline
  & & $Z_{4,t+h}$ & &  & & $Z_{5,t+h}$  & & & &   $Z_{6,t+h}$&  \\
 \hline
    $h$ & 1 & 2 & 3 & 4 & 1 & 2 & 3 & 4 & 1 &2 &3   \\           
 \hline
1& 0 & 0 & 0.001 & 0.999 & - & 0 & 0.001 & 0.999 & 0.006 & 0.842 & 0.152\\
2& 0 & 0 & 0.002 & 0.998 & - & 0 & 0.002 & 0.998 & 0.007 &0.725 & 0.268\\
3& 0 & 0 & 0.003 & 0.997 & - & 0 & 0.003 & 0.997 &0.007 & 0.639 & 0.354\\
4& 0 & 0 &0.005 & 0.995 & - & 0 & 0.004 & 0.996 & 0.008 & 0.574 & 0.418\\
5& 0 & 0 & 0.009 & 0.992 & - & 0 & 0.004 & 0.996 & 0.008 & 0.528 & 0.464\\
6& 0 & 0 & 0.01 & 0.99 & -   & 0 & 0.005 & 0.955 & 0.008 & 0.489 & 0.503\\
7& 0 & 0 & 0.011 & 0.989 & - & 0 & 0.006 & 0.994 & 0.008 & 0.461 & 0.531\\
8& 0 & 0 & 0.013 & 0.987 & - & 0 & 0.007 & 0.993 & 0.007 & 0.442 & 0.550\\
9& 0 & 0 & 0.015 & 0.985 & - & 0 & 0.006 & 0.994 & 0.008 & 0.430 & 0.562\\
10&0 & 0 & 0.016 & 0.984 & - &  0 & 0.007 & 0.993 &0.007 & 0.420 & 0.573\\
11&0 & 0 & 0.017 & 0.983 & - & 0 & 0.007 & 0.993 &0.007 & 0.411 & 0.582\\ 
12&0 & 0 & 0.018 & 0.981 & - & 0 & 0.008 & 0.992 & 0.007 & 0.403 & 0.591\\
 \hline
\end{tabular}
\caption{\label{Tab. Forc_ProbsA} Relative frequency table of states based $10000$ simulations for each time series for Method A.}
\end{table}

\begin{table}
\setlength{\tabcolsep}{2pt}
\centering
\begin{tabular}{|c|cc|cc|cc|cc|cc|cc|} 
 \hline
    $h$ &  $\hat{Z}_{1,t+h}$\rule{0pt}{3ex}  & True  &  $\hat{Z}_{2,t+h}$\rule{0pt}{3ex}  & True   &  $\hat{Z}_{3,t+h}$\rule{0pt}{3ex}  & True   &  $\hat{Z}_{4,t+h}$\rule{0pt}{3ex}  & True   &  $\hat{Z}_{5,t+h}$\rule{0pt}{3ex}  & True   &  $\hat{Z}_{6,t+h}$\rule{0pt}{3ex}  & True \\           
 \hline
1 & 1 & 1 & 1 & 1 & 1 & 2 & 4 & 4 & 4 & 4 & 2 & 3\\
2 & 1 & 1 & 1 & 1 & 1 & 2 & 4 & 4 & 4 & 4 & 2 & 2\\ 
3 & 1 & 1 & 1 & 1 & 1 & 1 & 4 & 3 & 4 & 4 & 2 & 2\\
4 & 1 & 1 & 1 & 1 & 1 & 1 & 4 & 3 & 4 & 3 & 2 & 2\\
5 & 1 & 1 & 1 & 1 & 1 & 1 & 4 & 3 & 4 & 3 & 2 & 2\\
6 & 1 & 1 & 1 & 1 & 1 & 1 & 4 & 3 & 4 & 3 & 3 & 2\\
7 & 1 & 1 & 1 & 1 & 1 & 1 & 4 & 3 & 4 & 3 & 3 & 1\\
8 & 1 & 1 & 1 & 1 & 1 & 1 & 4 & 3 & 4 & 3 & 3 & 2\\
9 & 1 & 1 & 1 & 1 & 1 & 1 & 4 & 3 & 4 & 3 & 3 & 2\\
10& 1 & 1 & 1 & 1 & 1 & 2 & 4 & 3 & 4 & 3 & 3 & 1\\
11& 1 & 1 & 1 & 1 & 1 & 1 & 4 & 3 & 4 & 3 & 3 & 1\\
12& 1 & 1 & 1 & 1 & 2 & 1 & 4 & 3 & 4 & 3 & 3 & 1\\
 \hline
\end{tabular}
\caption{\label{Tab. Forc_StatesA} Forecasts based on $10000$ simulations and true values of the time series for Method A.}
\end{table}

\begin{table}
\centering
\begin{tabular}{|c|cccc|cccc|ccc|} 
\hline 
  & & $Z_{1,t+h}$ & &  & & $Z_{2,t+h}$  & & & &   $Z_{3,t+h}$&  \\
 \hline
    $h$ & 1 & 2 & 3 & 4 & 1 & 2 & 3 & 4 & 1 &2 &3   \\           
 \hline
1& 0.902 & 0.097 & 0.001 & 0 & 0.908 & 0.090 & 0.003 & 0 & 0.762 & 0.236 & 0.002 \\
2& 0.844 & 0.143 & 0.013 & 0 & 0.842 & 0.139 & 0.018 & 0.001 & 0.645 & 0.349 & 0.005\\
3& 0.808 & 0.165 & 0.026 & 0 & 0.783 & 0.179 & 0.034 & 0.004 & 0.595 & 0.394 & 0.011\\
4& 0.784 & 0.172 & 0.042 & 0.002 & 0.740 & 0.199 & 0.051 & 0.010 & 0.556 & 0.429 & 0.015\\
5& 0.758 & 0.182 & 0.057 & 0.003 & 0.708 & 0.204 & 0.070 & 0.018 & 0.556 & 0.424 & 0.020\\
6& 0.740 & 0.186 & 0.068 & 0.006 & 0.677 & 0.211 & 0.084 & 0.028 & 0.555 & 0.419 & 0.026\\
7& 0.726 & 0.188 & 0.079 & 0.007 & 0.644 & 0.220 & 0.094 & 0.041 & 0.533 & 0.438 & 0.029\\
8& 0.707 & 0.191 & 0.093 & 0.009 & 0.619 & 0.222 & 0.108 & 0.051 & 0.533 & 0.432 & 0.035\\
9& 0.692 & 0.193 & 0.103 & 0.011 & 0.598 & 0.222 & 0.117 & 0.063 & 0.527 & 0.434 & 0.039\\
10& 0.676 & 0.198 & 0.112 & 0.014 & 0.584 & 0.215 & 0.125 & 0.076 & 0.523 & 0.436 & 0.041\\
11& 0.663 & 0.200 & 0.122 & 0.014 & 0.563 & 0.218 & 0.131 & 0.088 & 0.529 & 0.429 & 0.042\\ 
12& 0.656 & 0.197 & 0.130 & 0.017 & 0.544 & 0.220 & 0.138 & 0.099 & 0.520 & 0.435 & 0.045\\
 \hline
  & & $Z_{4,t+h}$ & &  & & $Z_{5,t+h}$  & & & &   $Z_{6,t+h}$&  \\
 \hline
    $h$ & 1 & 2 & 3 & 4 & 1 & 2 & 3 & 4 & 1 &2 &3   \\           
 \hline
1& 0 & 0 & 0.039 & 0.961 & - & 0 & 0.040 & 0.959 & 0.045 & 0.675 & 0.280\\
2& 0 & 0.001 & 0.085 & 0.914 & - & 0.001 & 0.070 & 0.928 & 0.061 & 0.540 & 0.398\\
3& 0 & 0.003 & 0.123 & 0.874 & - & 0.003 & 0.084 & 0.913 & 0.072 & 0.486 & 0.442\\
4& 0 & 0.008 & 0.150 & 0.843 & - & 0.005 & 0.095 & 0.900 & 0.075 & 0.469 & 0.455\\
5& 0 & 0.010 & 0.183 & 0.807 & - & 0.008 & 0.105 & 0.887 & 0.088 & 0.452 & 0.460\\
6& 0 & 0.016 & 0.200 & 0.784 & - & 0.011 & 0.112 & 0.877 & 0.090 & 0.446 & 0.464\\
7& 0 & 0.018 & 0.209 & 0.772 & - & 0.013 & 0.120 & 0.867 & 0.095 & 0.442 & 0.463\\
8& 0 & 0.024 & 0.222 & 0.754 & - & 0.017 & 0.127 & 0.856 & 0.103 & 0.434 & 0.462\\
9& 0.001 & 0.029 & 0.230 & 0.740 & - & 0.019 & 0.136 & 0.845 & 0.106 & 0.435 & 0.459\\
10&0.001 & 0.037 & 0.232 & 0.731 & - & 0.024 &  0.141 & 0.835 & 0.109 & 0.435 & 0.454\\
11& 0.001 & 0.041 & 0.241 & 0.717 & - & 0.027 & 0.147 & 0.826 & 0.113 & 0.437 & 0.449\\ 
12& 0.002 & 0.048 & 0.245 & 0.705 & - & 0.029 & 0.155 & 0.817 & 0.118 & 0.431 & 0.451\\
 \hline
\end{tabular}
\caption{\label{Tab. Forc_ProbsB} Relative frequency table of states based $10000$ simulations for each time series for Method B.}
\end{table}

\begin{table}
\setlength{\tabcolsep}{2pt}
\centering
\begin{tabular}{|c|cc|cc|cc|cc|cc|cc|} 
 \hline
    $h$ &  $\hat{Z}_{1,t+h}$\rule{0pt}{3ex}  & True  &  $\hat{Z}_{2,t+h}$\rule{0pt}{3ex}  & True   &  $\hat{Z}_{3,t+h}$\rule{0pt}{3ex}  & True   &  $\hat{Z}_{4,t+h}$\rule{0pt}{3ex}  & True   &  $\hat{Z}_{5,t+h}$\rule{0pt}{3ex}  & True   &  $\hat{Z}_{6,t+h}$\rule{0pt}{3ex}  & True \\           
 \hline
1 & 1 & 1 & 1 & 1 & 1 & 2 & 4 & 4 & 4 & 4 & 2 & 3\\
2 & 1 & 1 & 1 & 1 & 1 & 2 & 4 & 4 & 4 & 4 & 2 & 2\\ 
3 & 1 & 1 & 1 & 1 & 1 & 1 & 4 & 3 & 4 & 4 & 2 & 2\\
4 & 1 & 1 & 1 & 1 & 1 & 1 & 4 & 3 & 4 & 3 & 2 & 2\\
5 & 1 & 1 & 1 & 1 & 1 & 1 & 4 & 3 & 4 & 3 & 3 & 2\\
6 & 1 & 1 & 1 & 1 & 1 & 1 & 4 & 3 & 4 & 3 & 3 & 2\\
7 & 1 & 1 & 1 & 1 & 1 & 1 & 4 & 3 & 4 & 3 & 3 & 1\\
8 & 1 & 1 & 1 & 1 & 1 & 1 & 4 & 3 & 4 & 3 & 3 & 2\\
9 & 1 & 1 & 1 & 1 & 1 & 1 & 4 & 3 & 4 & 3 & 3 & 2\\
10& 1 & 1 & 1 & 1 & 1 & 2 & 4 & 3 & 4 & 3 & 3 & 1\\
11& 1 & 1 & 1 & 1 & 1 & 1 & 4 & 3 & 4 & 3 & 3 & 1\\
12& 1 & 1 & 1 & 1 & 1 & 1 & 4 & 3 & 4 & 3 & 3 & 1\\
 \hline
\end{tabular}
\caption{\label{Tab. Forc_StatesB} Forecasts based on $10000$ simulations and true values of the time series for Method B}
\end{table}

\section{Conclusions}
\label{sec6}
In this work, we introduce a framework for jointly modelling multiple ordinal time series that captures both serial dependence within each series and cross-dependence between different series. The proposed approach specifies a univariate time series model for each variable and employs copulas to construct the multivariate dependence structure. To address the computational challenges associated with high-dimensional models, we adopt a pairwise likelihood approach that relies exclusively on bivariate models.

Model estimation is carried out using a two-stage procedure for maximising the pairwise likelihood. In the first stage, the log-likelihoods of all possible bivariate models are maximised independently. In the second stage, an aggregation step is applied to ensure unique parameter estimates across the system. This estimation strategy substantially reduces computational cost without compromising estimation accuracy. In particular, we propose a weighted averaging scheme in the second stage, where the weights are inversely related to the standard errors of the individual estimates. Simulation results indicate that the proposed method performs well across different sample sizes, and that the weighted average consistently yields slightly better performance than a simple average; we therefore recommend its use in practice.

The use of pairwise likelihood introduces additional challenges for forecasting. Since inference is based solely on information from bivariate models, the marginal model for each time series is obtained from multiple pairwise specifications. For forecasting individual series, we propose two alternative approaches. The first constructs a state forecast from each relevant bivariate model and then applies a majority rule to obtain the final forecast. The second averages the conditional state probabilities obtained from the corresponding bivariate models and selects as the final forecast the state with the highest aggregated probability.

There are several things to consider for extending the current model. The inclusion of covariates in the marginal models is straightforward and therefore has not been explicitly addressed in this paper. Although we employ GLM-based marginal models, these may be readily replaced by alternative approaches for modelling univariate ordinal time series. Moreover, different marginal models may be specified for each series, if this is more appropriate for capturing the distinct dynamics of individual processes.
With regard to cross-correlation, we assume the same copula family across all pairs in both the simulation study and the empirical application. This assumption has an appealing interpretation, as the bivariate marginals can be viewed as arising from an underlying but unknown multivariate model, with our approach serving as an approximation to that model. Nevertheless, this restriction is not essential. Allowing different copula families for different pairs of time series is a natural and flexible extension of the framework, though it gives rise to an interesting and non-trivial model selection problem.
Finally, higher order models are also possible at the cost of added complexity. Model selection, including the order selection can be  challenging.

\section*{Funding}
The research work was supported by the Hellenic Foundation 
for Research and Innovation (HFRI) under the 5th Call for 
HFRI PhD Fellowships (Fellowship Number: 20535.)

\bibliographystyle{apalike}
\bibliography{biblio}

\begin{thebibliography}{}

\bibitem[B{\"u}hlmann and Wyner, 1999]{buhlmann1999variable}
B{\"u}hlmann, P. and Wyner, A.~J. (1999).
\newblock Variable length {M}arkov chains.
\newblock {\em The Annals of Statistics}, 27(2):480--513.

\bibitem[Chaubert et~al., 2008]{chaubert2008multivariate}
Chaubert, F., Mortier, F., and Saint~Andr{\'e}, L. (2008).
\newblock Multivariate dynamic model for ordinal outcomes.
\newblock {\em Journal of Multivariate Analysis}, 99(8):1717--1732.

\bibitem[Cox and Reid, 2004]{cox2004note}
Cox, D.~R. and Reid, N. (2004).
\newblock A note on pseudolikelihood constructed from marginal densities.
\newblock {\em Biometrika}, 91(3):729--737.

\bibitem[Fieuws and Verbeke, 2006]{fieuws2006pairwise}
Fieuws, S. and Verbeke, G. (2006).
\newblock Pairwise fitting of mixed models for the joint modeling of multivariate longitudinal profiles.
\newblock {\em Biometrics}, 62(2):424--431.

\bibitem[Fokianos and Kedem, 2003]{fokianos2003regression}
Fokianos, K. and Kedem, B. (2003).
\newblock Regression theory for categorical time series.
\newblock {\em Statistical Science}, 18(3):357--376.

\bibitem[G{\"o}ttlein and Pruscha, 1992]{gottlein1992ordinal}
G{\"o}ttlein, A. and Pruscha, H. (1992).
\newblock Ordinal time series models with application to forest damage data.
\newblock In {\em Advances in GLIM and Statistical Modelling: Proceedings of the GLIM92 Conference and the 7th International Workshop on Statistical Modelling, Munich, 13--17 July 1992}, pages 113--118. Springer.

\bibitem[Hirk et~al., 2019]{hirk2019multivariate}
Hirk, R., Hornik, K., and Vana, L. (2019).
\newblock Multivariate ordinal regression models: an analysis of corporate credit ratings.
\newblock {\em Statistical Methods \& Applications}, 28:507--539.

\bibitem[Hirk et~al., 2022]{hirk2022corporate}
Hirk, R., Vana, L., and Hornik, K. (2022).
\newblock A corporate credit rating model with autoregressive errors.
\newblock {\em Journal of Empirical Finance}, 69:224--240.

\bibitem[Hui et~al., 2018]{hui2018sparse}
Hui, F.~K., M{\"u}ller, S., and Welsh, A. (2018).
\newblock Sparse pairwise likelihood estimation for multivariate longitudinal mixed models.
\newblock {\em Journal of the American Statistical Association}, 113(524):1759--1769.

\bibitem[Jahn and Wei{\ss}, 2024]{jahn2024nonlinear}
Jahn, M. and Wei{\ss}, C.~H. (2024).
\newblock Nonlinear {GARCH}-type models for ordinal time series.
\newblock {\em Stochastic Environmental Research and Risk Assessment}, 38(2):637--649.

\bibitem[Jahn and Wei{\ss}, 2025]{jahn2025modeling}
Jahn, M. and Wei{\ss}, C.~H. (2025).
\newblock Modeling multivariate ordinal time series.
\newblock {\em Journal of Applied Statistics}, pages 1--28.

\bibitem[Kazianka and Pilz, 2010]{kazianka2010copula}
Kazianka, H. and Pilz, J. (2010).
\newblock Copula-based geostatistical modeling of continuous and discrete data including covariates.
\newblock {\em Stochastic Environmental Research and Risk Assessment}, 24(5):661--673.

\bibitem[Liu et~al., 2022a]{liu2022modeling2}
Liu, M., Li, Q., and Zhu, F. (2022a).
\newblock Modeling air quality level with a flexible categorical autoregression.
\newblock {\em Stochastic Environmental Research and Risk Assessment}, pages 1--11.

\bibitem[Liu et~al., 2022b]{liu2022modeling1}
Liu, M., Zhu, F., and Zhu, K. (2022b).
\newblock Modeling normalcy-dominant ordinal time series: an application to air quality level.
\newblock {\em Journal of Time Series Analysis}, 43(3):460--478.

\bibitem[Nalpantidi and Karlis, 2025]{nalpantidi2025bivariate}
Nalpantidi, A. and Karlis, D. (2025).
\newblock A bivariate dar ($1 $) model for ordinal time series.
\newblock {\em arXiv preprint arXiv:2510.05680}.

\bibitem[Nikoloulopoulos and Mentzakis, 2016]{nikoloulopoulos2016copula}
Nikoloulopoulos, A.~K. and Mentzakis, E. (2016).
\newblock A copula-based model for multivariate ordinal panel data: application to well-being composition.
\newblock {\em arXiv preprint arXiv:1604.05643}.

\bibitem[Nikoloulopoulos and Moffatt, 2019]{nikoloulopoulos2019coupling}
Nikoloulopoulos, A.~K. and Moffatt, P.~G. (2019).
\newblock Coupling couples with copulas: analysis of assortative matching on risk attitude.
\newblock {\em Economic Inquiry}, 57(1):654--666.

\bibitem[Pruscha, 1993]{pruscha1993categorical}
Pruscha, H. (1993).
\newblock Categorical time semes with a recursive scheme and with covariates.
\newblock {\em Statistics}, 24(1):43--57.

\bibitem[Raftery, 1985]{raftery1985model}
Raftery, A.~E. (1985).
\newblock A model for high-order {M}arkov chains.
\newblock {\em Journal of the Royal Statistical Society Series B: Statistical Methodology}, 47(3):528--539.

\bibitem[Vana-G{\"u}r, 2024]{vana2024multivariate}
Vana-G{\"u}r, L. (2024).
\newblock Multivariate ordinal regression for multiple repeated measurements.
\newblock {\em arXiv preprint arXiv:2402.00610}.

\bibitem[Varin et~al., 2011]{varin2011overview}
Varin, C., Reid, N., and Firth, D. (2011).
\newblock An overview of composite likelihood methods.
\newblock {\em Statistica Sinica}, pages 5--42.

\bibitem[Varin and Vidoni, 2006]{varin2006pairwise}
Varin, C. and Vidoni, P. (2006).
\newblock Pairwise likelihood inference for ordinal categorical time series.
\newblock {\em Computational statistics \& data analysis}, 51(4):2365--2373.

\bibitem[Wang et~al., 2023]{wang2023forecast}
Wang, X., Hyndman, R.~J., Li, F., and Kang, Y. (2023).
\newblock Forecast combinations: An over 50-year review.
\newblock {\em International Journal of Forecasting}, 39(4):1518--1547.

\bibitem[Wei{\ss}, 2019]{weiss2019distance}
Wei{\ss}, C.~H. (2019).
\newblock Distance-based analysis of ordinal data and ordinal time series.
\newblock {\em Journal of the American Statistical Association}.

\bibitem[Wei{\ss}, 2023]{weiss2023ordinal}
Wei{\ss}, C.~H. (2023).
\newblock Ordinal compositional data and time series.
\newblock {\em Statistical Modelling}, page 1471082X231190971.

\bibitem[Wei{\ss} and G{\"o}b, 2008]{weiss2008measuring}
Wei{\ss}, C.~H. and G{\"o}b, R. (2008).
\newblock Measuring serial dependence in categorical time series.
\newblock {\em AStA Advances in Statistical Analysis}, 92:71--89.

\end{thebibliography}

\section*{APPENDIX}

\subsection*{Comparing estimation methods}
We would like to compare three methods of estimation: the full maximum likelihood (FL),  the pairwise likelihood (PL) given in  (\ref{Eq: Pairwise log-likelihood}) and the two-stage algorithm for pairwise likelihood (TS) used in the paper. To make FL approach feasible, we are using dimension $d=3$ where the full likelihood is possible since it involves only evaluation of a trivariate copula. The purpose of this simulation is to show that the two-step method has minor efficiency loss relative to the (intractable) FL and it is equivalent to the PL.

We have generated $B=100$ samples of sample size $T=100$ and $T=500$ from a trivariate ordinal time series. Each time series $Z_{kt}$, $k=1,2,3$ and $t=1,\ldots,T$ takes values in $\mathcal{S}=\{1,2,3\}$. Their joint trivariate distribution is described by a Gaussian copula:

\begin{align*}
    C_{R}(u_1,u_2,u_3)=\Phi_R(\Phi^{-1}(u_1),\Phi^{-1}(u_2),\Phi^{-1}(u_3)), 
\end{align*}
where $\Phi(.)$ the inverse cumulative distribution function of a standard normal  and $\Phi_R(.)$ the joint cumulative distribution function of a trivariate normal distribution with mean vector zero and covariance matrix equal to the correlation matrix $R$:
$$R=\begin{bmatrix}
    1& \rho_{12} & \rho_{13}\\ 
    \rho_{12} & 1 & \rho_{23}\\
    \rho_{13} & \rho_{23} & 1\\
\end{bmatrix}.$$

Each time series marginally follows an ordinal autoregressive model of order $1$ including also cross-correlation terms. Namely, 

$$Z_{kt}|\mathcal{F}_{t-1} \sim Multinomial(1;\pi^{(k)}_{1t},\pi^{(k)}_{2t},\pi^{(k)}_{3t})$$
\begin{align*}
\logit(\gamma^{(k)}_{jt})=\alpha^{(k)}_{0j}+\alpha_{k1}Z_{1,t-1}+\alpha_{k2}Z_{2,t-1}+\alpha_{k3}Z_{3,t-1}
\end{align*}
for $k=1,2,3,~~$ $j=1,2,~~$ $t=2,\ldots,T$. $\mathcal{F}_{t-1}=\bm{\sigma}\{{Z_{1i},Z_{2i},Z_{3i},i\leq t-1}\}$ is a $\sigma-$algebra containing all the history of the system and $\gamma^{(k)}_{jt}=P(Z_{kt}\leq j|\mathcal{F}_{t-1})$, for $k=1,2,3$ and $j=1,2$. The true parameters are given below: 
\begin{align*}
&~\alpha^{(1)}_{01}=-0.5,\alpha^{(1)}_{02}=0.5,\alpha_{11}=0.4,\alpha_{12}=0.2,\alpha_{13}=0.15, \\
&~\alpha^{(2)}_{01}=-0.3,\alpha^{(2)}_{02}=0.7,\alpha_{21}=0.15,\alpha_{22}=0.35,\alpha_{23}=0.2,\\
&~\alpha^{(3)}_{01}=-0.4,\alpha^{(3)}_{02}=0.8,\alpha_{31}=0.3,\alpha_{32}=0.2,\alpha_{33}=0.45\\
&~\rho_{12}=0.5,\rho_{13}=-0.3, \rho_{23}=0.2
\end{align*}

\begin{figure}
    \centering
    \includegraphics[width=0.9\linewidth]{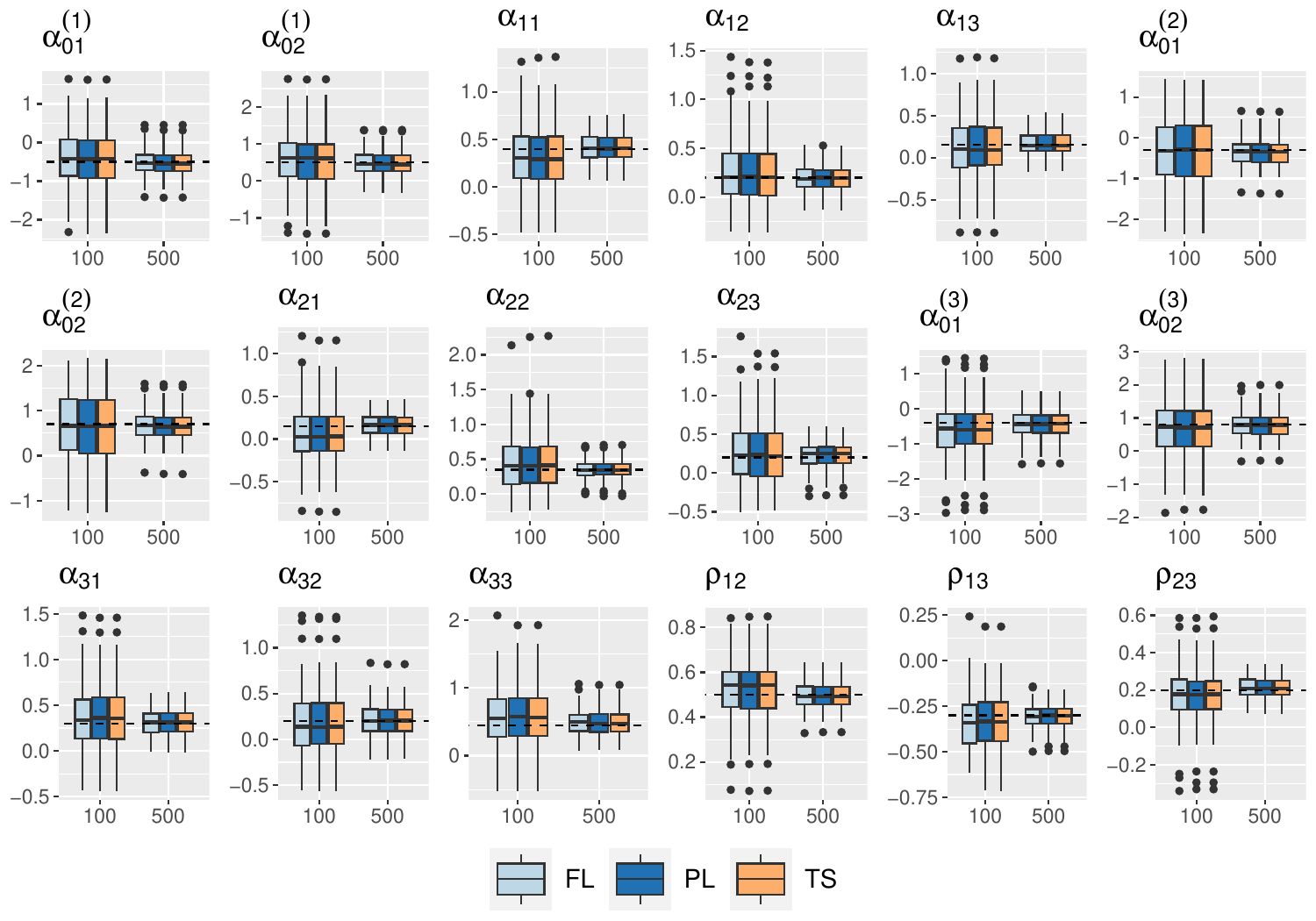}
    \caption{Estimates per method of estimation and sample size.}
    \label{Methods_Comparison}
\end{figure}

Figure \ref{Methods_Comparison} presents boxplots of the parameter estimates obtained using the three methods across different sample sizes. One can see that the two step approach shows the same performance with the other methods without any clear difference.

\begin{table}[H]
\centering
\begin{tabular}{ccc}
& $\mbox{var}_{FL}/\mbox{var}_{TS}$ & $\mbox{var}_{PL}/\mbox{var}_{TS}$ \\
\hline
$n=100$ & 0.996 & 0.995 \\
$n=500$ & 0.998 & 0.999 \\
\hline
\end{tabular}
\caption{\label{eff} Relative efficiency of the three methods. FL is the full likelihood, PL the pairwise likelihood 
maximized alltogether and TS the proposed method that maximizes separately for each pair and then synthesizes the results. 
}
\end{table}

We have also calculated for each parameter, the ratio of the variance of the method relative to that of the FL approach. Averaging over the 16 parameters we get the efficiency. Those numbers can be seen in Table \ref{eff}.
The first column shows the case when the FL is compared to the TS method. FL shows negligible gain but recall that FL is not available as dimension increases. 
The second column compares the TS with the PL. The numbers show that there is not a real difference. 
This makes clear that the loss of efficiency is minimal. 

As far as the computing time, 
the two-step method was  1.65 times faster than PL in both $n=100$ and $n=500$. A direct comparison with the FL is not fair, since the implementation is very different, but the PL and TS methods were hugely faster (the TS was about 50 times faster than the FL). For the FL in the $d=3$ dimensions, one needs to evaluate trivariate Gaussian integrals which are by far more demanding. For bivariate Gaussian integrals the vectorized library \texttt{pbivnorm} can speed a lot the computations. 
Also note that in the $3$-dimensional example the TS solves 3 maximization problems with 6 parameters, while the PL  solves one problem with 18 parameters which is much time consuming. The difference in higher dimensions is more pronounced, keeping in mind the difficulty sometimes to estimate so many parameters. Finally, recall that FL is not available for larger dimensions so there is a clear evidence that the two-step approach can provide quite quick and efficient estimates in all cases contrary to PL and FL.

\end{document}